\documentclass[11pt]{article}
\usepackage{graphicx}

\usepackage{fontawesome} 
\usepackage{algorithm}
\usepackage{algpseudocode}
\usepackage{jheppub}
\usepackage{hyperref}
\usepackage{amsmath,amsfonts,amssymb,appendix,comment}
\usepackage[usenames,dvipsnames,table]{xcolor}
\usepackage{slashed,subfigure,setspace,footnote,multirow,longtable,braket}
\usepackage{epsfig,mathrsfs,latexsym,color,url,etoolbox}
\usepackage{caption}
\usepackage{bbm}
\definecolor{nicered}{rgb}{0.7,0.1,0.1}
\definecolor{nicegreen}{rgb}{0.1,0.5,0.1}
\definecolor{CarnationPink}{rgb}{1.0, 0.65, 0.79}
\DeclareMathAlphabet{\mathbbold}{U}{bbold}{m}{n}    

\usepackage{comment}
\usepackage{xspace}
\usepackage{slashed}
\usepackage{float}

\usepackage{siunitx}
\allowdisplaybreaks

\usepackage{physics}
\usepackage{dcolumn}
\usepackage{bm}
\usepackage[table]{xcolor}
\usepackage{multirow}
\usepackage{comment}
\usepackage{url}
\usepackage{tcolorbox}
\usepackage[capitalize]{cleveref}
\usepackage[T1]{fontenc}
\usepackage{cases}

\newcommand{\beq}{\begin{equation}}
\newcommand{\eeq}{\end{equation}}
\def\calchep{\texttt{CalcHEP}\xspace}
\def\PETITE{\texttt{PETITE}\xspace}
\def\vegas{\texttt{VEGAS}\xspace}
\def\GEANTfour{\texttt{GEANT-4}\xspace}
\def\EGS{\texttt{EGS-5}\xspace}
\def\FLUKA{\texttt{FLUKA}\xspace}
\def\absQ{Q}
\def \kappaCoup{\varepsilon}

\newcommand{\GeV}{\mathrm{GeV}}
\newcommand{\MeV}{\mathrm{MeV}}

\DeclareMathOperator{\sgn}{sgn}
\def\madgraph{\texttt{MadGraph}\xspace}

\def\maddump{\texttt{MadDump}\xspace}
\newcommand{\e}{{\rm e}}
\newcommand{\iu}{{\rm i}}
\newcommand{\lambdabar}{{\mkern0.75mu\mathchar '26\mkern -9.75mu\lambda}}

\crefname{section}{Sec.}{Secs.}
\crefname{figure}{Fig.}{Figs.}
\crefname{equation}{Eq.}{Eqs.}
\crefname{table}{Table}{Tables}

\definecolor{nicegreen}{rgb}{0., 0.75, 0.46}
\definecolor{azure}{rgb}{0.97, 1.0, 1.0}

\begin{document}

\preprint{CALT-TH/2024-001, FERMILAB-PUB-24-0003-T, CERN-TH-2024-006, MI-HET-825}

\title{Dark fluxes from electromagnetic cascades  {{\href{https://github.com/kjkellyphys/PETITE}{\large\color{BlueViolet}\faGithub}}  }  }
 

\author[1,2]{Nikita Blinov,}
\author[3]{Patrick J. Fox,}
\author[4,5]{Kevin J. Kelly,}
\author[3]{Pedro A.N. Machado,}
\author[6]{Ryan Plestid}
\affiliation[1]{Department of Physics and Astronomy, University of Victoria, Victoria, BC V8P 5C2, Canada}
\affiliation[2]{Department of Physics and Astronomy, York University, Toronto, Ontario, M3J 1P3, Canada}
\affiliation[3]{Particle Theory Department, Fermi National Accelerator Laboratory, Batavia, IL 60510, USA}
\affiliation[4]{Theoretical Physics Department, CERN, Esplande des Particules, 1211 Geneva 23, Switzerland}
\affiliation[5]{Department of Physics and Astronomy, Mitchell Institute for Fundamental Physics and Astronomy, Texas A\&M University, College Station, TX 77843, USA}
\affiliation[6]{Walter Burke Institute for Theoretical Physics, California Institute of Technology, Pasadena, CA 91125, USA}
\emailAdd{nblinov@yorku.ca}
\emailAdd{pjfox@fnal.gov}
\emailAdd{kjkelly@tamu.edu}
\emailAdd{pmachado@fnal.gov}
\emailAdd{rplestid@caltech.edu}

\date{\today}

\abstract{We study dark sector production in electromagnetic (EM) cascades.  
This problem requires accurate simulations of Standard Model (SM) and dark sector processes, both of which impact angular and energy distributions of emitted particles that ultimately determine flux predictions in a downstream detector.
We describe the minimal set of QED processes which must be included to faithfully reproduce a SM cascade, and identify a universal algorithm to generate a dark sector flux given a Monte-Carlo simulation of a SM shower. We provide a new tool, \PETITE, which simulates EM cascades with associated dark vector production, and compare it against existing literature and ``off the shelf'' tools. 
The signal predictions at downstream detectors can strongly depend on the nontrivial interplay (and modelling) of SM and dark sector processes, in particular multiple Coulomb scattering and positron annihilation. We comment on potential impacts of these effects for realistic experimental setups.
}

\maketitle
\pagebreak

\section{Introduction \label{sec:Introduction-new} }

General ultraviolet (UV) completions of the Standard Model (SM)~\cite{Craig:2022cef}, the existence of dark matter~\cite{Bertone:2004pz, Battaglieri:2017aum}, and neutrino masses~\cite{Kajita:2016cak, McDonald:2016ixn} motivate the existence of feebly interacting particles \cite{Beacham:2019nyx,Antel:2023hkf}. 
If the mass scale of these particles is below the electroweak scale, they must be singlets under the SM gauge group~\cite{L3:2001xsz, CMS:2019hsm, ATLAS:2023sbu}, dramatically restricting the possible set of interactions with the dark sector relevant at experimentally-accessible energies. 
This observation has lead to several benchmark ``portal'' interactions to be identified as the primary targets for experimental investigation~\cite{Beacham:2019nyx, Gori:2022vri, Harris:2022vnx}.

 Intensity frontier experiments with thin or thick targets~\cite{Berlin:2018bsc, NA64:2023wbi, deNiverville:2011it, Batell:2014mga, SHiP:2015vad, BDX:2016akw, deGouvea:2018cfv, DeRomeri:2019kic, Machado:2019oxb, Harnik:2019zee, ArgoNeuT:2019ckq, DUNE:2020fgq, Breitbach:2021gvv, Apyan:2022tsd, Coloma:2023adi} often provide the most sensitive probes of dark sectors, especially for particle masses below a few GeV~\cite{Beacham:2019nyx, Gori:2022vri}. 
 It is therefore important to quantify and maximize the impact of near term intensity frontier searches.
 Crucially, this requires a detailed understanding of both the production and detection of physics beyond the SM (BSM). 
 Visible detector signatures involve
 an energy deposition in a detector via a decay into visible particles or via scattering. 
 The former is easily evaluated, while the latter requires some atomic and/or nuclear physics input. For the purposes of dark sector searches a theoretical precision of $\mathcal{O}(10\%)$ is acceptable, and this is achievable in many instances.

The determination of production of BSM particles in a thick target/beam stop is more challenging. 
High energy particles generate hadronic and electromagnetic (EM) cascades within the target leading to an exponentially growing multiplicity of final states with energies ranging over multiple decades \cite{Bhabha:1937xxx,Gaisser:1990vg}.
The production of dark states from these cascades has a complicated dependence upon the nature of the couplings to the dark sector, and on the distribution of energies and particle multiplicities in the hadronic and electromagnetic sectors.  Particularly for low-mass dark states, there are many possible sources of emission such as meson decay and bremsstrahlung.
Historically this complexity has been circumvented in one of two ways. First, one can consider BSM particle production only from the attenuated primary beam, since the attenuation can be estimated analytically for lepton beams~\cite{Tsai:1966js}. Alternatively, one can focus on production in the first interaction length, such that attenuation and effects of subsequent hadronic and EM cascades can be neglected~\cite{deNiverville:2011it,deNiverville:2016rqh,Magill:2018tbb,Magill:2018jla,Berryman:2019dme,Plestid:2020kdm}. 
Both approaches are conservative, systematically {\it under-predicting} the dark sector production rates.  
This guarantees that excluded regions of parameter space based on total rate analyses are robust against uncertainties, since additional production channels can only strengthen the bounds obtained from null observations. 

For electrophilic gauge extensions of the SM, e.g., $L_e-L_\mu$ or $L_e-L_\tau$, the dominant production mode for dark states is from the EM shower. Similarly, in certain axion models secondary electrons dominate axion production even in a proton beam dump~\cite{Tsai:1986tx}.  Conversely, if a dark photon couples democratically to charged leptons and hadrons, it has large production rates from meson decays~\cite{Batell:2009di,Reece:2009un} or proton bremsstrahlung~\cite{Blumlein:2013cua,Foroughi-Abari:2021zbm}.  While we do not specialize to a particular UV model, we will focus on the former case where the EM shower is the dominant source of dark sector particles.
Thus, in this work we study 
EM shower evolution and its contributions to dark sector particle production in thick target environments. 
While this problem has been discussed previously \cite{Buonocore:2018xjk,Nardi:2018cxi,Celentano:2020vtu,Capozzi:2021nmp}, a variety of different approximations have been used in the literature, some of which are mutually incompatible; there is currently no consensus on which predictions are reliable. 
With this in mind, our goals in this work are the following:
\begin{enumerate}
    \item To provide a (SM) generator-independent solution for producing a dark sector flux from a SM cascade. 
    \item To describe and implement a principled, flexible, and systematically improvable framework for producing a Monte Carlo (MC) event record of dark particles emanating from EM showers. 
    \item To compare existing results on BSM particles produced in EM showers, and to identify which methods are reliable and to diagnose the cause of any disagreements.
\end{enumerate}
EM showers are largely driven by a small set of reactions all of which are perturbatively calculable and systematically improvable if needed. We adopt the phenomenologically-motivated assumption that the impact of dark sector emissions on the dynamics of EM showers is negligible.  
This allows us to ``factorize'' the SM cascade from dark sector production, enabling a self-contained treatment of each. 
The problem of dark sector production in an EM shower therefore represents a {\it perturbatively calculable} problem that can be computed via a series of controllable approximations.
In fact, a similar approach could also be used for dark sector production in hadronic showers, but we do not explore this possibility here.

To distill the EM dynamics down to their minimal components, we have implemented a custom lightweight package, \PETITE ({\bf P}ackage for {\bf E}lectromagnetic {\bf T}ransitions {\bf I}n {\bf T}hick-target {\bf E}nvironments), specialized for the production of light dark sector particles (in addition to electrons, photons and positrons) in a beam-dump or other fixed target experiment with ${\sim}1{-}100$~GeV beams.\!\footnote{Strictly speaking the approximations we use are valid for electron, positron, and photon energies greater than $\sim 1~{\rm MeV}$, and less than $100~{\rm GeV}$ where the Landau-Pomeranchuk-Migdal effect substantially modifies bremsstrahlung cross sections \cite{Landau:1953um,Migdal:1956tc}.}
We make use of well-documented and controlled approximations whenever possible, and model matter as a set of screened Coulomb sources, and a homogeneous gas of electrons at rest. 
We find that a good description of SM dynamics can be obtained with just three hard processes and a reasonable model of both continuous energy losses and multiple Coulomb scattering. 
We will show that some of these central processes are mismodelled in existing SM simulation tools. 
We also identify disagreements between existing results in the literature for BSM production in EM cascades, stemming from practical or conceptual issues.  
For example, at the practical level, stable numerical simulation of light particle bremsstrahlung requires careful MC sampling of phase space, which is difficult to achieve with many existing MC tools due to approximate phase space singularities. 
At the conceptual level, we find that certain methods for reprocessing a SM simulation of an EM cascade to produce dark sector states based on rescaling of cross sections are theoretically inconsistent. This method predicts incorrect kinematics of BSM particles, and can lead to order-of-magnitude errors in the signal rate for realistic experimental setups.
Our simulation tool, \PETITE, does not attempt to replicate or supplant existing tools such as \GEANTfour~\cite{GEANT4:2002zbu,Allison:2006ve,Allison:2016lfl}, however it can serve as a lightweight, fast MC and/or as a useful diagnostic tool. 
Important questions can be asked, and  answered, using \PETITE such as
\begin{itemize}
    \item How do SM uncertainties on cascade propagation influence BSM fluxes? 
    \item How should one efficiently simulate and/or sample BSM events from a SM shower?
    \item Which methodologies are reliable? Do they agree? And if not, why?
\end{itemize}
Although far from exhaustive, we try to answer an interesting subset of these questions in \cref{Pheno}. We focus on five concrete phenomenological issues which are relevant for dark sector production in EM showers: 1) the impact of multiple Coulomb scattering (MCS) modelling on geometric acceptances, 2) the treatment of resonant dark vector production and its interplay with continuous energy loss and MCS, 3) kinematic correlations in simulations of SM bremsstrahlung, 
4) a comparison of dark sector bremsstrahlung in \PETITE  to an approximate method that instead re-purposes simulated SM bremsstrahlung kinematics, 
and 5) calculation of total fluxes at downstream detectors for a set of realistic experimental configurations, in a full \PETITE simulation.

The goal of this paper is to provide a brief, but pedagogical, introduction to the physics of EM cascades, and to explain how BSM microphysics should be consistently implemented in any future simulation. It is organized as follows. In \cref{SM-cascades} we describe EM cascades in the SM and their implementation within \PETITE. In \cref{BSM-cascades} we discuss relevant production mechanisms for dark sector states, specializing to massive vector bosons. Our focus is on EM production, however we also highlight for the reader regions of parameter space in which hadronic production is known to dominate the flux in UV-complete models like the dark photon. In \cref{Pheno} we investigate the phenomenological questions listed above. In each case we focus on experimentally motivated distributions, but defer detailed analyses of specific experiments to future work. 
Finally in \cref{Conclusions} we summarize our findings and comment on future directions.

We discuss several technical details in the appendices. 
In \cref{A:Landau} we collect the known expressions for QED processes needed for simulating EM cascades and discuss various approximations that were used to obtain them. 
In \cref{A:formfactors} we describe the atomic and nuclear form factors that enter dark and SM bremsstrahlung, and pair production. 
The \PETITE models for continuous energy loss and MCS are described in \cref{A:continous_processes}. 
Fast and reliable simulation of SM processes necessitated a multitude of variable transformations and integrator tuning which we discuss in \cref{A:vegas}. In \cref{A:SM_shower_validation} we validate the implementation of these processes by comparing \PETITE to a semi-analytic shower model and to \GEANTfour. In \cref{A:darkbremm} we describe the numerical integration of dark sector bremsstrahlung and compare our implementation to other available programs. Finally, we provide explicit expressions for dark Compton scattering, and $e^+ e^-\rightarrow \gamma V$ in \cref{A:darkcompton}.
Our implementation of positron annihilation into dark vectors includes soft- and collinear photon resummation which is discussed in \cref{A:radiative-return}. 

\section{Electromagnetic cascades \label{SM-cascades}}

We first discuss the processes necessary to describe EM cascades involving electrons, positrons, and photons, before explaining how they are modelled in \PETITE. These particles can interact with atomic electrons or nuclei in the material.  In nuclear interactions, we treat the atoms as static and include a form factor which accounts for both atomic screening and the loss of nuclear coherence as the momentum exchanged with the target varies. These form factors are described in \cref{A:formfactors}. 
In what follows we denote nuclei by their atomic charge $Z$.
Here we focus on the dominant components of EM showers, namely $e^\pm$ and $\gamma$, but muons can be included straightforwardly. 
\subsection{Relevant processes \label{subsec:sm_processes}}
There are five processes which dominate the dynamics of an electromagnetic cascade.  These can be classified as ``discrete'' and quasi-continuous.
For discrete processes we generate kinematics of outgoing particles in individual events -- expressions for the differential cross sections can be found in \cref{A:Landau}.  
The three reactions of this type are
\begin{enumerate}
    \item \textbf{Bremsstrahlung on atomic nuclei: $e^\pm Z \rightarrow e^\pm Z \gamma$.} 
    This process is dominated by small momentum transfers, $Q$, that satisfy $Q R\ll1$ where 
    $R$ is the nuclear radius. We treat the nucleus as a static fixed  Coulomb potential, use the Bethe-Heitler matrix element \cite{Bethe:1934za}, and include 
    form factors to account for screening and finite nuclear size effects at small and large momentum transfers respectively \cite{Tsai:1973py}, see \cref{A:formfactors}.   
    A minimum energy cut (whose default value is $\omega_{\rm cut} = 1\, \mathrm{MeV}$), is applied to avoid infrared (IR) 
    singularities from soft photons, 
    see \cref{A:Landau} for further details. 
    While this cut determines the number of discrete ``steps'' taken in a cascade, BSM observables are not sensitive to this as long as the dark sector particle mass or detector threshold is larger than $\omega_{\rm cut}$. 
    \item \textbf{Bethe-Heitler pair production: $\gamma Z \rightarrow e^+e^- Z$.} 
    The matrix element for pair production is related to that of bremsstrahlung via crossing
    symmetry. The kinematics are similar, and we make use of the same small angle approximations, 
    Thomas-Fermi screening form factor, and Bethe-Heitler matrix element. The process is IR finite
    and no minimum energy cut is required. 
    \item \textbf{Scattering on atomic electrons :  $(\gamma,e^\pm) ~e^-_{\rm bound} \rightarrow (\gamma,e^\pm) e^-$ and $e^+ e^-_{\rm bound}\rightarrow \gamma \gamma$.} 
     We use tree-level expressions for Bhabha, M{\o}ller, and Compton scattering 
     as well as for $e^+ e^-$ annihilation, see \crefrange{eq:compton}{eq:moller}. Atomic electrons are modelled as free and at rest.  These reactions tend to dominate at lower energies and are responsible for an 
     asymmetry between electron and positron populations. Atomic electrons that are ionized
     above threshold in the simulation lead to an excess of $e^-$ over $e^+$.  
     For these processes, we enforce a minimum energy $E_{\rm min}$ of the outgoing $e^\pm$.
\end{enumerate}
In addition to these reactions, there are two processes that are much more common when $e^\pm, \gamma$ transit material.
They are so frequent that they must be modelled as quasi-continuous in practice; we will refer to these collisions as ``soft'' to contrast them with hard reactions above which are discrete in nature. The soft processes we consider are
\begin{enumerate}
    \item \textbf{Coulomb scattering on nuclei: $e^\pm Z \rightarrow e^\pm Z$.} 
    This process is dominated by frequent small angle scattering which leads to a Gaussian
    angular distribution with power law tails that can be modelled using multiple scattering theory \cite{Bethe:1953va,Lynch:1990sq}.  We discuss the impact of different MCS models on dark sector fluxes in \cref{sec:MCS} (the specific models implemented in \PETITE are described in \cref{A:multiple-scatter}) and highlight the importance of MCS in positron annihilation in \cref{subsec:rad-return}.
    \item \textbf{Ionization: $e^\pm  e^-_{\rm bound} \rightarrow e^\pm  e^-_{\rm free} $.}
    Atomic ionization dominates energy loss for electrons at energies below the critical energy, $\mathcal{O}(10-100\;\MeV)$ depending on $Z$~\cite{ParticleDataGroup:2022pth}.  Furthermore, at higher energies where bremsstrahlung is more important (and which we simulate explicitly) ionization can still influence the shower dynamics. We treat ionization energy losses 
    with a constant $\dd E/\dd x = 2 ~{\rm MeV} ~{\rm cm}^2/{\rm g}$, i.e., independent of energy (this number is only weakly dependent on the type of atom~\cite{ParticleDataGroup:2022pth}). 
\end{enumerate}
In addition to these two processes there are contributions to energy loss coming from soft photons emitted in bremsstrahlung where the outgoing photon energy is below $\omega_{\mathrm{cut}}$.  However, for the default value of the cut that separates hard from soft,  $\omega_{\mathrm{cut}}=1\,\MeV$, the contribution from below-threshold bremsstrahlung will be subdominant to the energy loss from ionization, see \cref{A:continous_processes} for more details.

The small number of processes, along with the applicability of perturbation theory makes EM cascades well suited to a systematic investigation of secondary production. We make further approximations for some of the reactions above that we now detail before moving on to discuss \PETITE itself.

Pair production is subject to collinear divergences, while bremsstrahlung has overlapping soft and collinear divergences. However a further singularity exists in both processes which actually dominates the phenomenology and this is the low-$Q^2$ singularity of Coulomb scattering. This makes the phase space for bremsstrahlung and pair production very peaked at small $Q^2$. We make use of small-angle approximations that treat the collinear singularities analytically.  At the same time this approach simplifies the treatment of the Coulomb singularity. The relevant expressions for differential cross-sections can be found in \cite{Berestetskii:1982qgu} and also in \cref{A:Landau}.

\subsection{Implementation within \PETITE}\label{subsec:implementation_of_petite}
We now focus on the explicit implementation of the above processes in \PETITE. The design of \PETITE involves a Python class called \texttt{Shower}. A \texttt{Shower} is initiated with a material (e.g.,\ graphite with $Z=6$), a minimum energy cut ($E_{\rm min}\ge \omega_{\rm cut}$) below which particles are not tracked dynamically, and a dictionary of auxiliary information (e.g.\ the differential cross sections as a function of energy). 
 
A \texttt{Shower} is a collection of particles produced in a cascade, each of which is represented by the \texttt{Particle} class.
This class contains all relevant information for a given particle, including its four-momentum, position, and a dictionary of additional information. 
This dictionary contains a unique ID number, the Particle Data Group identifying number (PDG-ID), its parent's PDG-ID and the process and particle from which it was generated, its generation number (the number of hard interactions before the particle was created), and weight (for weighted events). 
Other info such as the mass and stability of the particle are also stored.

An EM cascade is generated by calling the \texttt{Shower} method \texttt{generate\_shower} which takes as an input an initial \texttt{Particle} that seeds the shower, e.g., an electron or a photon with four momentum $(E,p_x,p_y,p_z)$ entering at position $(0,0,0)$.
By instantiating a \texttt{Shower} with arguments describing the initial particle, material, $E_{\rm min}$, and path of pre-computed \texttt{VEGAS} integrators, \PETITE sets up the  necessary infrastructure to calculate the showers. 
The shower itself is generated by calling \texttt{generate\_shower}, which loops through all \texttt{Particle} in the \texttt{Shower}. At each iteration, \texttt{Particles} are propagated using the method \texttt{propagate\_particle} which determines the step size $x_{\rm step}$ and whether a discrete/hard interaction takes place in that step; the probability of such an interaction is $P=1-{\rm e}^{-x_{\rm step}/\lambda(E)}$ with $\lambda(E)$ the mean free path calculated using the tabulated cross-sections stored in \texttt{Shower}.\footnote{The step size $x_{\rm step} \approx \lambda(E)/10$ is chosen small enough such that $\lambda(E)$ does not change significantly over the step, and large enough such that the probability of interaction is not too small.} The four-momentum and position of the \texttt{Particle} are updated at each step, taking into account multiple scattering and energy loss (see above and \cref{A:energy-loss}).
Steps are repeated until a reaction takes place (as determined by an accept-reject algorithm using the probability $P$), or until the energy of the particle falls below $E_{\rm min}$. 

Having conditioned on a reaction occurring, the reaction type is randomly selected according to the relative cross-sections (``branching ratios'') available to the current \texttt{Particle}. The functions used in the shower generation that encode the physical processes (e.g., bremsstrahlung, Compton scattering, etc), can be found in \texttt{all\_processes.py} and \texttt{radiative\_return.py}, while MCS implementations are found in \texttt{moliere.py}.  The most computationally intensive step is generating the final state kinematics for the selected process. 
In \PETITE this is achieved using importance sampling from a pre-trained \vegas integrator at a nearby (slightly larger) energy chosen such that the phase space of the pre-trained integrator is a strict superset of the phase space of interest, see~\cref{A:vegas}. The nearby energy is used as an estimator, but the matrix element used for importance sampling is evaluated at the energy of interest. This enables us to correctly (and relatively quickly) sample the full multiparticle phase space in each reaction. 
The \texttt{generate\_shower} loop continues until the energy of all active particles has dropped below $E_{\rm min}$.  Note that while \PETITE comes with pre-computed \texttt{VEGAS} integrators, the user may want to generate others, for example a coarser grid of generators or new integrators for different dark vector masses.
To this end, PETITE comes with a set of utilities to facilitate such a task.

Parent particles are kept in the event record, along with their daughters which are created at every vertex. 
For example, after a bremsstrahlung event $e^+ Z \rightarrow e^+\gamma Z$ the event record will contain the initial parent positron $e^+$, the daughter positron $e^+$, and the daughter photon $\gamma$. All the information of the parent positron is kept for bookkeeping, but it is not propagated further. Daughter particles propagate until interacting.
Thus, each \texttt{Shower} contains the full shower history, and the user can then plot distributions, and/or event displays using the provided utilities. 

\begin{figure}[t]
    \centering
    \includegraphics[width=\linewidth]{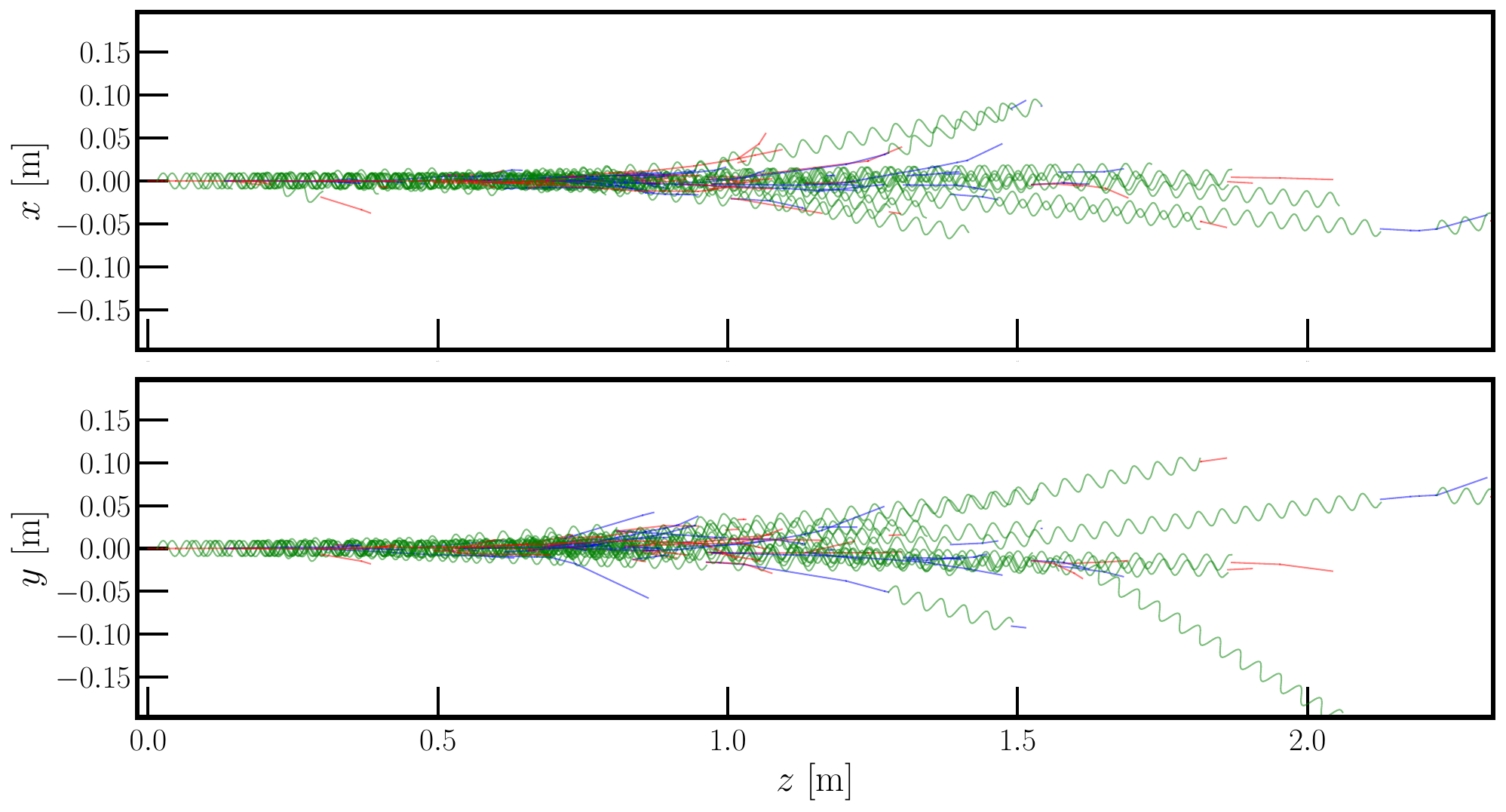}
    \caption{The projection in $x-z$ (top) and $y-z$ (bottom), where $z$ is along the beam axis, of a typical shower development as simulated using \PETITE.  In this case the shower is initiated by a $10~{\rm GeV}$ photon incident on a thick graphite target. Photons are shown as green wavy lines, electrons as blue straight lines, and positrons as red straight lines.  Particles have been tracked until their energy drops below 3~MeV.}
    \label{fig:event_display}
\end{figure}

A code-focused guide for \PETITE, along with some example applications, is available on the associated GitHub repository \href{https://github.com/kjkellyphys/PETITE}{\large\color{BlueViolet}\faGithub}.
A visualization of the output from \PETITE is shown in \cref{fig:event_display}. A comparison of the performance of \PETITE and \GEANTfour is provided in \cref{A:SM_shower_validation}.

\section{Dark sector fluxes from electromagnetic cascades \label{BSM-cascades}}
We now  focus on the flux of dark sector particles that can emerge from an EM cascade in a thick target. 
Some of the earliest discussions of light particle production in beam dumps occurred in the context of searches for fractional charges~\cite{Bellamy:1967qf}, heavy leptons~\cite{Barna:1968su} and neutrino-like particles~\cite{SLAC:E56,Rothenberg:1972pt}. In these experimental works, analytic approximations of the secondary spectra of photons and electrons, due to Tsai and Whitis~\cite{Tsai:1966js}, were used to estimate signal yields in thick targets.
Later Tsai's analytic results were applied to axion bremsstrahlung from an attenuated electron beam~\cite{Donnelly:1978ty,Tsai:1986tx}. This method has been adopted to reinterpret old experimental results in the context of other dark sector particles like the dark photon~\cite{Bjorken:2009mm,Andreas:2012mt}. 
A different signal prediction technique was used in analyzing the electron beam dump E137 data~\cite{Bjorken:1988as}: simulated SM shower spectra were convolved with cross-sections for multiple axion production mechanisms, including bremsstrahlung, annihilation, and Primakoff-type reactions. To our knowledge this was one of the first works to systematically consider a full set of processes possible in a EM shower for a dark sector model. In Ref.~\cite{Marsicano:2018vin,Rella:2022len} the secondary production of muons is considered since the focus of those works is on muonphillic interactions. 
More recently in Refs.~\cite{Nardi:2018cxi,Celentano:2020vtu}, it was realized that for generic dark sector couplings, resonant $e^+e^- \rightarrow X$ production can be extremely efficient for $m_X \lesssim 50~{\rm MeV}$, possibly supplying the dominant flux of new physics particles at experiments with primary or secondary positrons. This qualitative conclusion was supported by an independent study in Ref.~\cite{Capozzi:2021nmp}. Secondary fluxes of photons have also been demonstrated to be of interest in the context of axion like particles \cite{Tsai:1986tx}. 

Recently, several approaches for beam dump dark sector production have been developed. In Ref.~\cite{Capozzi:2021nmp} a scheme was proposed to model dark vector fluxes by reweighing the distribution of SM photons produced in a \GEANTfour simulation; crucially this procedure does not account for possible differences in angular distributions of visible and dark states. 
Two \GEANTfour native solutions to the production of dark sector states in medium were presented in Refs.~\cite{Bondi:2021nfp,Eichlersmith:2022bit}. 
In these works, SM cascade and dark sector production are simulated simultaneously by using an unphysically large dark sector coupling. As emphasized in Ref.~\cite{Celentano:2020vtu}, this approach is inefficient in two distinct ways. First, the dark sector coupling cannot be made too large; otherwise the properties of the SM cascades are altered, ultimately leading to inaccurate predictions of dark sector fluxes. The second disadvantage is that one cannot reuse the expensive part of the simulation (the SM cascade) as one scans over BSM particle masses and portal types. The authors of Ref.~\cite{Celentano:2020vtu} have implemented a solution to the above problems by separately simulating the SM cascade in \GEANTfour and BSM production in \maddump~\cite{Buonocore:2018xjk} or \texttt{BdNMC}~\cite{deNiverville:2016rqh}. By linking multiple independent codes, it becomes difficult to assess systematic uncertainties in SM or dark sector production modelling.

Our goal is to develop an algorithm to predict dark sector production from all processes that can occur in a given model, which can address some of the issues highlighted above.  
As a concrete example we will study the case of a dark sector vector $V$ with mass $m_V$ which couples to predominantly to the electron vector current.  There are a multitude of UV-complete models which contain this, and other couplings, in their low energy limit.  In general, at the energies we are concerned with, the IR couplings could be to baryons, leptons, and mesons:
\begin{equation}
\mathcal{L}\supset V_\mu\left(c_e  \bar e \gamma^\mu e + c_p \bar p \gamma^\mu p + c_n \bar n \gamma^\mu n \right) + c_\pi \frac{e}{32\pi^2 f_\pi} \pi^0\epsilon^{\mu\nu\alpha\beta} F_{\mu\nu} \partial_\alpha V_\beta + \ldots
    \label{eq:interaction}
\end{equation}
In this work we will primarily focus on the case where the coupling to electrons is the dominant coupling, $c_e \equiv \kappaCoup e$ and $c_{p,n,\pi}\approx 0$.  This can occur for electrophilic gauge extensions of the SM, e.g., $L_e-L_\mu$ or $L_e-L_\tau$, and EM shower-induced production dominates. Similarly, in certain axion models secondary electrons dominate axion production even in a proton beam dump~\cite{Tsai:1986tx}, although through a different Lorentz structure.  Alternatively, a dark photon couples democratically to charged leptons and hadrons ($|c_p|=|c_e|= c_\pi =\kappaCoup e$), giving large production rates from meson decays~\cite{Batell:2009di,Reece:2009un} or proton bremsstrahlung~\cite{Blumlein:2013cua,Foroughi-Abari:2021zbm}. 
We will not explicitly discuss hadronic production modes and refer the interested reader to the relevant literature, e.g.,~\cite{deNiverville:2011it,deNiverville:2016rqh,Magill:2018tbb,Magill:2018jla,Berryman:2019dme,Plestid:2020kdm}.

\subsection{Beyond the Standard Model fluxes from electromagnetic cascades \label{BSM-general} }

EM production in a thick target stemming from~\cref{eq:interaction} is dominated by three main mechanisms: 
\begin{enumerate}
    \item \textbf{Dark bremsstrahlung: $e^\pm Z \rightarrow e^\pm Z V$.} 
    In the static limit (i.e., the nucleus is taken as infinitely heavy), this process is kinematically allowed for $E_e\geq m_V$. Due to the non-zero dark vector mass, the matrix element for dark bremsstrahlung contains collinear, but not soft divergences. The collinear divergences only appear in the limit where both the electron and vector boson can be treated as approximately massless. This results in a matrix element that prefers $E_V\gg m_V$.  In turn this means dark bremsstrahlung dominantly produces hard dark vectors~\cite{Bjorken:2009mm,Liu:2016mqv,Liu:2017htz,Tsai:1989vw,Gninenko:2017yus}.  This highly peaked differential cross-section presents challenges for fast and accurate MC sampling.  In \cref{A:darkbremm} we discuss our approach, using the adaptive \vegas algorithm~\cite{Lepage:1977sw,Lepage:2020tgj} to sample the differential cross-sections presented in~\cite{Tsai:1989vw, Gninenko:2017yus}.
    \item \textbf{Dark Compton scattering: $\gamma  e^-_{\rm bound} \rightarrow V e^-_{\rm free} $.} 
    This process is operational provided $ E_\gamma \geq m_V^2/(2 m_e)$. The corresponding differential cross section can be found in~\cref{A:darkcompton}.
    \item \textbf{Positron annihilation: $e^+  e^-_{\rm bound} \rightarrow V(n\gamma) $.} 
    This channel can be the largest contribution to the dark vector flux when it is open. 
    The kinematic threshold is set by $s=s_{\rm th}$ with $s_{\rm th } = m_V^2$. 
    For $s> s_{\rm th}$ the emission of initial state radiation of soft and collinear photons 
    brings the energy onto the resonant peak. This is the radiative return effect seen in $e^+e^-$ collisions. 
    At threshold the tree-level expression for $e^+e^- \rightarrow V \gamma$ diverges like $\sigma \propto 1/(s-s_{\rm th})$. When integrated over an incoming positron energy distribution, this produces a logarithmic singularity. One-loop radiative corrections are need to ensure the cross section is finite. This issue can be dealt with at tree level by using an IR cutoff on the outgoing photon energy \cite{Celentano:2020vtu}. Here we follow the lepton-collider literature \cite{Kuraev:1985hb,Nicrosini:1986sm,Karliner:2015tga,Greco:2016izi} and make use of resummed QED distribution functions for the electrons and positrons in the collision. This removes the singularity in the vicinity of the reaction threshold, and resums large double-logarithms (see \cref{subsec:rad-return} and~\cref{A:radiative-return} for more details).
\end{enumerate}

\subsection{Generating dark vectors from a \texttt{Shower} using \PETITE \label{BSM-PETITE} \label{algorithm} }

We now describe a model-independent algorithm for the production of new weakly-coupled physics in a SM cascade.
The small coupling between SM and the dark sector ensures that production of dark states is a small perturbation on the evolution of the shower.
Thus, the dark production and the SM shower development can be simulated separately at leading non-trivial order in the small coupling.
While we focus on the specific implementation within \PETITE, the general set-up outlined here can be reproduced in other codes. 
Importantly, because the dark sector production is entirely separate from the generation of the SM cascade, any standard shower simulation tool can be used to generate the initial SM shower. This can then be ``dressed'' with BSM production.

Whether generated by \PETITE, or in another simulation, let us assume that we have a list of SM particle momenta, $\vec{p}$, and production positions, $\vec{x}$, associated with a SM shower. 
Our goal is then to produce a possible BSM interaction position and probability for each particle in this list. 
For each dark sector process $p$ available to the SM particle, the probability of interaction $\dd w^{(p)}$ 
occurring between $z$ and $z + \dd z$ is proportional to the probability for the BSM process to occur and to the probability of no SM process happening up to that point: 

\begin{equation}\label{eq:dweightdz}
    \frac{\dd w^{(p)}}{\dd z} =  n^{(p)} 
    \sigma^{(p)}_{\rm BSM}(z) ~\exp\left(-\int_{0}^{z} \frac{\dd z'}{\lambda_{\rm MFP}^{\rm SM}(z)}\right), 
\end{equation}
where $z=0$ corresponds to the SM particle's starting position $\vec{x}$. The cross section $\sigma^{(p)}_{\rm BSM}$ and mean free path $\lambda_{\rm MFP}^{\rm SM}$ depend on $z$ through the particle's energy $E(z)$; for photons no continuous energy loss occurs and so the energy is fixed. We assume a homogeneous target with number density $n^{(p)}$, which is process-dependent. Note that the mean free path is determined by SM processes in the small dark vector coupling limit, i.e., $\lambda_{\rm MFP} \approx \lambda_{\rm MFP}^{\rm SM}$. Sampling $z$ from this distribution therefore gives an interaction energy $E(z)$ at which the differential cross-section for $p$ is sampled to generate dark vector kinematics. This event is assigned the weight $w^{(p)} = \int \dd z\; \dd w^{(p)}/\dd z$. The collection of dark sector particle momenta produced in this way represents a set of possible shower histories which produced a single dark sector particle each. 
 
For most processes the distribution in \cref{eq:dweightdz} is relatively flat, but, as discussed above, the cross section for positron annihilation, $e^+e^-\to V(n\gamma)$, is highly energy dependent with a peak at $s=m_V^2$. 
Due to the multiple Coulomb scattering, there can be significant deflection of the parent particle between its initial energy $E_0=E(z=0)$ and the energy at which the dark vector is produced. It is therefore crucial for resonant $e^+e^- \rightarrow V(n\gamma)$ production to properly sample from \cref{eq:dweightdz}, \cref{subsec:rad-return} highlights the impact of this sampling procedure.

The \PETITE algorithm for producing dark states from SM cascade is summarized below. 
\begin{algorithm}[!h]
   \caption*{\PETITE algorithm}
    \begin{algorithmic}[100]
    \State  $shower = $ existing Standard Model shower (e.g.~as generated by \PETITE)
\For{$particle$ in $shower$}    
\If{$particle = e^+$}
    \State $processlist = \{\texttt{Dark\  bremsstrahlung, Dark annihilation}\}$  
\ElsIf{$particle = e^-$}
    \State $processlist = \{\texttt{Dark\  bremsstrahlung}\}$
\ElsIf{$particle = \gamma$}
    \State $processlist = \{\texttt{Dark\ Compton}\}$
\EndIf
\For{$p$ in $processlist$}
\If{energy of $particle$ is above energy threshold for process $p$}
    \State  \textbullet ~determine the interaction position/energy by sampling from \cref{eq:dweightdz}. 
    \State \textbullet ~generate dark sector kinematics of the outgoing BSM particle by drawing from $\dd\sigma_{\rm BSM}^{(p)}$.
    \State \textbullet ~compute weight $w^{(p)}$ by integrating \cref{eq:dweightdz}. 
    \State \textbullet ~record the weight $w^{(p)}$ position, momentum, and particle species.
\EndIf
\EndFor
\EndFor
\end{algorithmic}
\end{algorithm}

\noindent We will occasionally refer to this procedure as resampling. It generates a list of weighted dark sector emissions that could occur in a SM shower and it can be repeated for several distinct SM showers to obtain average predictions for dark sector production. 
Note that the procedure outlined above produces an equal number of weighted events for each available dark process. Processes can instead be sampled based on their relative importance within \PETITE. 

This algorithm is implemented in \PETITE in the \texttt{DarkShower} class. 
A \texttt{DarkShower} is initiated by passing it a \texttt{Shower} object which contains a pre-simulated SM shower. 
This can be performed either by specifying an external file location or passing a \texttt{Shower} object created by \PETITE. 
~The BSM events are produced by running the method \texttt{generate\_dark\_shower}, which loops through all particles in the shower producing weighted events.

\subsection{Algorithm for long cascades \label{long-cascade}}
In the discussion above we have assumed that less than one dark sector particle is produced per electromagnetic shower. The multiplicity of particles in a shower grows rapidly with  incident energy, which  can result in $\mathcal{O}({\rm few})$ or greater dark sector particles being produced within a single shower. In this case one must take care to account for the correlations among the descendants of a given parent particle. For instance, suppose our parent particle is an electron with energy $E_0$. Let us suppose that the probability of dark sector emission can be estimated by $P({\rm DS } | e^-) \sim \mathcal{O}(\kappaCoup^2)$. If the typical number of SM daughters in a shower is $\langle N_{\rm SM}\rangle$ and 
\begin{equation} 
   \kappaCoup^2 \times \langle N_{\rm SM}\rangle  \sim \mathcal{O}(1) , 
\end{equation}
then one must consider multiple dark emissions within a single progenitor’s shower. To do this one would draw the number of daughters from a Poisson distribution with the appropriate mean, then sample each daughter sequentially, altering the subsequent dynamics of the SM shower along each leg. 

If $\langle N_D \rangle$ is so large that the ultra-weak coupling limit is not valid, then the algorithm discussed above can be viewed as a subroutine. In this more general case one first draws the number of dark sector particles produced, executes \PETITE algorithm, then modifies the shower dynamics along the appropriate leg. The procedure is repeated until the $\mathcal{O}({\rm few})$ dark sector events are produced after which the SM simulation is set back to its initial configuration and the entire procedure repeated.

\subsection{Comparison to  previous work \label{Litt-Rev}}
\PETITE differs from previous implementations discussed in the introduction of this section in four key ways. First, having a minimal SM reaction network, using \PETITE we have been able to pinpoint the impact of SM simulation details on downstream fluxes. In practice, this means that we can easily turn off different SM processes, modify their modelling, and study their effects on dark sector yields. We provide some examples of these studies in~\cref{Pheno}. Second, in contrast to Refs.~\cite{Bondi:2021nfp,Eichlersmith:2022bit} we can re-purpose the same SM shower for many different BSM parameter choices allowing for efficient parameter scans. Third, in contrast to Ref.~\cite{Capozzi:2021nmp} we sample BSM events from their true underlying differential cross section. This avoids issues that stem from very different differential distributions of SM and BSM bremsstrahlung. This problem is explored in more detail in~\cref{subsec:sm_vs_ds_brem}. Finally, in contrast to Ref.~\cite{Celentano:2020vtu} we work at the level of MC event record, rather than energy-angle distributions. This is advantageous for tracking uncertainties and allows greater flexibility for interfacing with different event generators. Implementation differences aside, we find reasonable, although not perfect, agreement of dark vector yields between \PETITE and Ref.~\cite{Celentano:2020vtu} in~\cref{subsec:yields}.

\section{Key processes and their impact on phenomenology \label{Pheno}}
We now discuss applications of \PETITE to experimentally-relevant questions, and compare its predictions to existing methods for dark particle production in a thick target. 
The typical experimental setup we are interested in consists of a beam impinging a thick target and a downstream detector at a small solid angle from the target, such as the MiniBooNE, DUNE and BDX experiments.
While specific detector signatures rely on the details of a given experimental setup, we focus on the predictions at the level of the dark sector particle flux. 
Several experiments of interest have a proton beam, which can initiate a EM cascade through $\pi^0$ production and decay, and through bremsstrahlung. 
Since \PETITE does not simulate these hadronic interactions, we obtain the EM showers  either from \GEANTfour, which simulates the full hadronic and EM showers, or from \texttt{PYTHIA}~\cite{Bierlich:2022pfr}, which provides the primary $\pi^0$ whose decay photons are showered with \PETITE. 
Overall, we find only marginal differences between \GEANTfour and \texttt{PYTHIA} when using them to predict relevant observables.

In what follows we consider both SM and BSM processes whose detailed implementation can potentially impact experimental results. First, we examine the modelling of multiple Coulomb scattering (MCS) and its effect on the angular spread of SM cascades and of the resulting dark vectors in \cref{sec:MCS}. MCS turns out to have an interesting interplay with dark sector production from resonant positron annihilation, which we discuss in~\cref{subsec:rad-return}.
Next we describe the challenges of simulating SM bremsstrahlung and compare \PETITE to other software in~\cref{subsec:sm_brem}; we find that the sampled phase space distributions can differ significantly between different codes, but these 
differences can be washed out in thick targets where, e.g., MCS plays an important role in determining the shower angular spread. In~\cref{subsec:sm_vs_ds_brem} we highlight the qualitative differences between SM and dark sector bremsstrahlung which make it difficult to map SM photons into dark sector particle emissions, as was done previously. 
Finally in \cref{subsec:yields} we use \PETITE to estimate dark sector signal yields for a realistic set of experimental parameters. 
See also \cref{A:SM_shower_validation} for  validations of \PETITE standard EM showers against \GEANTfour.

\subsection{Multiple scattering and downstream fluxes \label{sec:MCS} }
There are a variety of Standard Model effects that can conceivably influence downstream fluxes.
We will start by studying how the modelling of MCS can change the angular distribution during a shower's development, thereby affecting downstream acceptances. 
Modelling differences can arise due to the specific implementation in a code (e.g., choices of particle step size) or variations in the microscopic input (e.g., the 
probability distribution of scattering angles). 
Here we will focus on the latter effect.

The goal of MCS theory is to compute the distribution of scattering angles $f(\theta,t)$ after a charged particle has traversed a thickness $t$ of material. Analytic calculations of this quantity rely on approximating the atomic potential by the Thomas-Fermi model~\cite{Moliere:1947zza,Bethe:1953va}. Comparisons of MCS models implemented in \GEANTfour find ${\sim}10\%$ agreement with empirical measurements \cite{Gottschalk:1993xxx,Makarova:2017xxx}. 
The general form of this distribution is Gaussian with power law tails, which result from the central limit of a large number of small angle deflections and rare large angle scatters, respectively. 
In many practical applications, only the Gaussian approximation is used with different choices of its width~\cite{Lynch:1990sq}.
We will illustrate the potential impact of MCS modelling by calculating the angular spread of dark vectors resulting from EM showers simulated with two different MCS models.

We start with the method of Moli\`ere \cite{Moliere:1947zza}, which was improved by Bethe \cite{Bethe:1953va}. 
We will summarize Bethe's results to explain the different approximations used in computing MCS.
The number of electrons in an angular interval $\dd\theta$ after traversing a material thickness $t$ is $f(\theta,t)\theta \dd\theta$.
Moli\`ere theory allows an asymptotic expansion of $f(\theta,t)$ in a parameter, $B$, which encodes the depth traversed in a given material. Specifically, $B$ is the solution of the transcendental equation 
\begin{equation}
  B-\ln B = \ln \frac{\chi_c^2}{1.167\chi_a^{2}},
  \label{eq:bethe_transcendental}
\end{equation}
where $\chi_c^2\equiv4\pi N t \alpha^2 Z(Z+1)z^2/(p\beta)^2$ is a reference scattering angle that characterizes the probability of Rutherford scattering, with  $N$ the atomic number density, $Z$ the atomic number, and $z$, $p$ and $\beta$ the charge, momentum and velocity of the propagating particle; the screening angle $\chi_a^{2}= (1.13+3.76\tilde\alpha^2)\lambdabar^2/a^2$ encodes the suppression of Rutherford scattering, where $\tilde\alpha=zZ\alpha/\beta$, and $\lambdabar=1/p$ is the de Broglie wavelength of the electron and $a=0.885 Z^{-1/3}/(\alpha m_e)$ is the Fermi radius of the atom. The screening angle can be determined by calculating the single scattering by a Thomas-Fermi potential. More details are given in Ref.~\cite{Bethe:1953va}, and in \cref{A:continous_processes}. The ratio $\chi_c^2/\chi_a^{2}$ is approximately the mean number of scatters in a thickness $t$~\cite{Bethe:1953va,Lynch:1990sq}, so $\chi_c^2/\chi_a^{2} \gg 1$ (so $B\sim \ln \chi_c^2/\chi_a^2$) unless $t$ is so small that the Moli\`ere theory would no longer be applicable. 
The expansion developed by Bethe can be summarized as
\begin{equation}
    \label{eq:MCS_bethe}
    f(\theta, t)= \frac{1}{\chi_c^2 B}\left[ f_0(\theta) + B^{-1} f_1(\theta) + B^{-2} f_2(\theta) + \ldots \right], 
\end{equation}
where $f_i$ are given as integrals containing Bessel functions.  In particular, $f_0$ is a Gaussian and the higher $f_i$ lead to a broadening of the tails of this Gaussian. 
The point here is that this approximation, while not trivial to compute, can be systematically improved, as Bethe argues that $B$ typically assumes values between 5 and 20, and the approximation can be valid even for large angles.

The second model we consider is a Gaussian-only approximation to MCS due to Lynch and Dahl~\cite{Lynch:1990sq}. The width of the Gaussian is also characterized by $\chi_c^2$ and $\chi_a^{2}$ as detailed in \cref{A:continous_processes}, with numerical factors fit to a Monte Carlo simulation of many discrete Coulomb scatters. This prescription is more accurate than the slightly simpler Gaussian distribution given in the PDG~\cite{ParticleDataGroup:2022pth}.

\begin{figure}[t]
    \centering  \includegraphics[height=0.345\linewidth,trim={1.1cm 0 0.5cm 0}, clip]{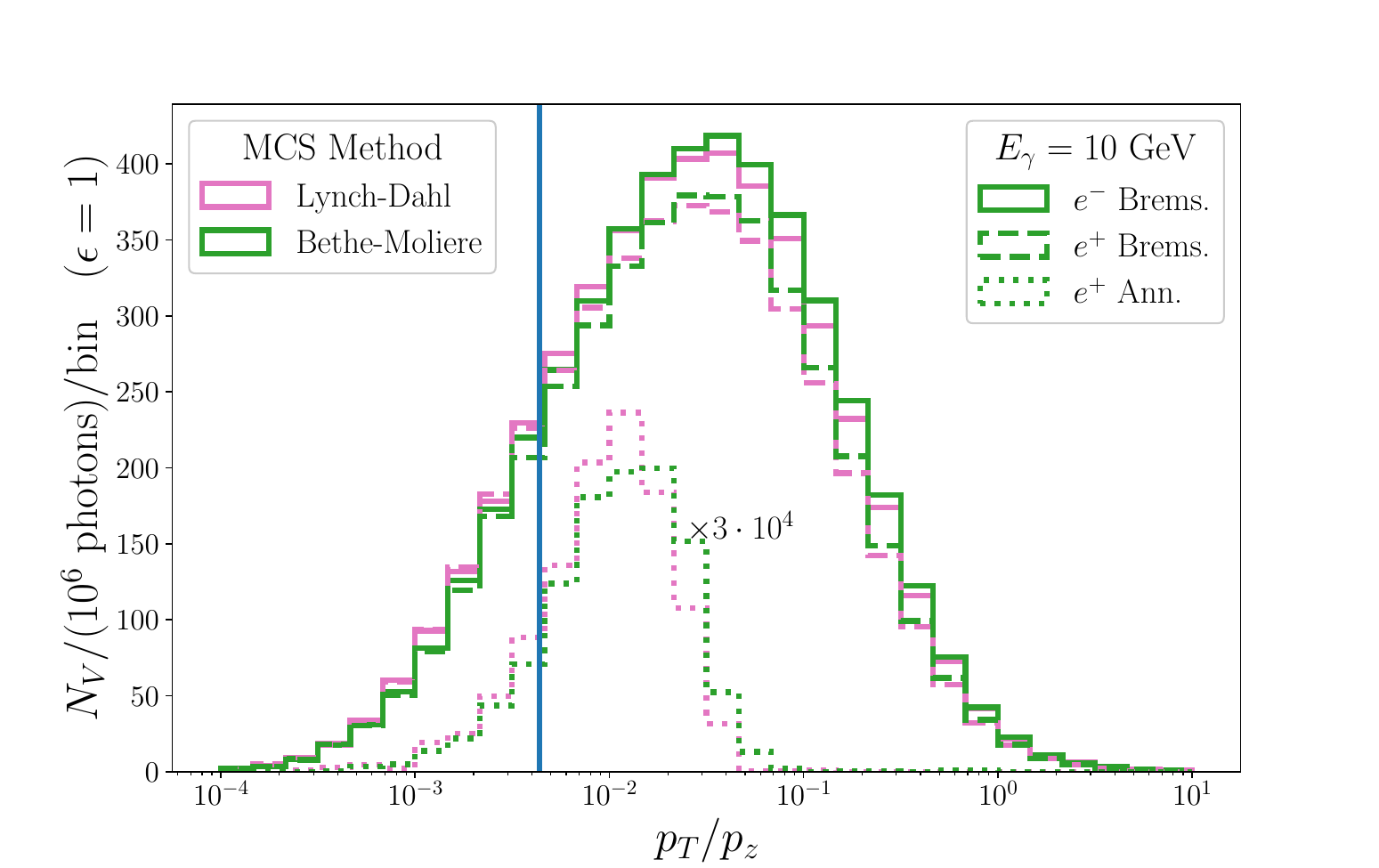}~\includegraphics[height=0.345\linewidth,trim={1.1cm 0 0.5cm 0}, clip]{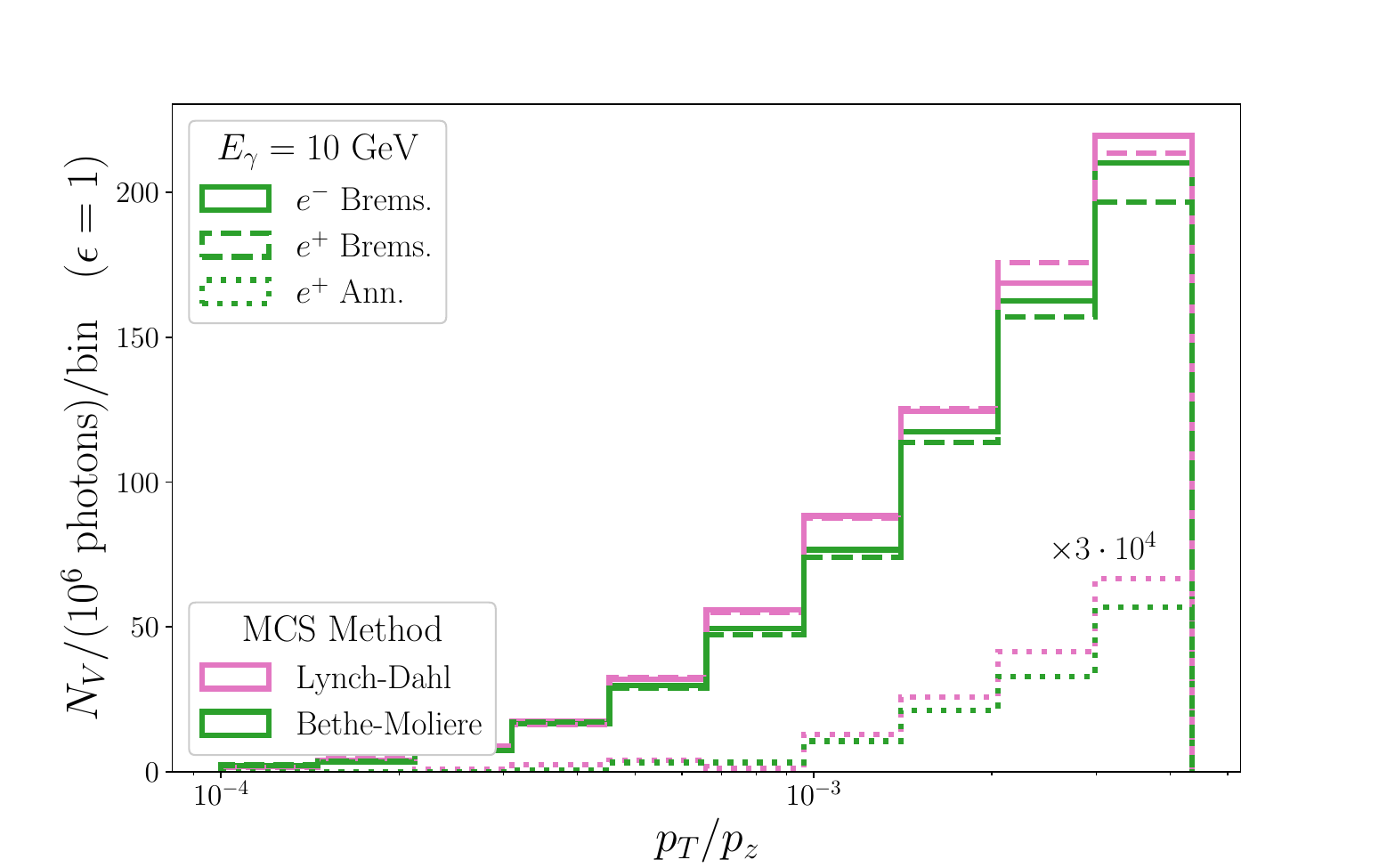}
   \caption{Impact of MCS on BSM fluxes for a $10~{\rm GeV}$ incident photon, on graphite, producing dark vectors with mass $m_V=5~{\rm MeV}$: (\textbf{Left}) Distribution of $p_T/p_z$ for dark bremsstrahlung from positrons and electrons, and dark annihilation of positrons on atomic electrons. Distributions are shown for two different MCS implementations. A vertical blue line shows a rough angular acceptance for the DUNE near detector $(2.5\;\mathrm{m}/574\;\mathrm{m}=4.3~\mathrm{mrad})$. Typical photon energies from $\pi^0$ decays from a 120 GeV proton beam range from 1-30 GeV. (\textbf{Right}) The same distribution below the cut $p_T/p_Z \leq 2.5/574$. Results are obtained from 3000 independent cascades simulated with \PETITE. The annihilation probability is roughly four orders of magnitude higher than the probability of dark bremsstrahlung and the curves have been re-scaled so as to fit on the same axes.  \label{fig:MCS-impact} }
\end{figure}

In the left panel of \cref{fig:MCS-impact} we show the impact of the choice of MCS model on the angular distribution of dark vectors of mass 5 MeV produced in EM showers initiated by 
a 10 GeV photon impinging a carbon target.  Note that the typical photon energy from the decay of $\pi^0$ produced when a $120$ GeV proton beam hits a target is $1-30$ GeV.
The green curve labelled ``Bethe-Moli\`ere'' is obtained by retaining the first two terms in the expansion discussed above, $f_0$ and $f_1$. The magenta curves correspond to the Gaussian Lynch-Dahl MCS model.  

To better appreciate the impact of MCS modelling, we show a DUNE-like acceptance as a vertical blue line ($2.5\;\mathrm{m}/574\;\mathrm{m}=4.3\;\mathrm{mrad}$), and present a ``zoomed in'' plot of the accepted events in the right panel. 
The more complete Bethe-Moli\`ere theory predicts a systematic shift to larger values of $p_T$, with effects on the order of ${\sim}10\%$, and can lead to some geometrical acceptance impact on downstream detectors. For example, the geometric acceptance for dark vectors with mass $m_V=5~{\rm MeV}$ produced via resonant annihilation changes from $\epsilon_{\rm geo}\approx 0.49$ when using the Lynch-Dahl implementation to $\epsilon_{\rm geo}\approx 0.43$ when using the more accurate Bethe-Moliere theory.

\subsection{Resonant positron annihilation \label{subsec:rad-return}}

Positron annihilation to dark sector states via $e^+e^-\rightarrow V (n\gamma)$ has been shown to be the most efficient production mechanism available in EM cascades for light dark sector masses $m_V \lesssim 100~{\rm MeV}$. 
This is true in a general EM cascade \cite{Nardi:2018cxi}, and even in proton beam dump experiments \cite{Celentano:2020vtu,Capozzi:2021nmp}. 
Recent results from the NA64 collaboration explicitly make use of resonant production \cite{Andreev:2021fzd,NA64:2023wbi}. 

Despite being the dominant production mechanism for certain dark vector masses, theoretical treatments of $e^+e^- \rightarrow V (n\gamma)$ in the literature do not account for several important effects.
For experiments with a detector far away from the target the most relevant effect is the interplay of annihilation and the modelling of MCS (the other effect, modelling of radiative return, is discussed in~\cref{A:radiative-return}).
Because the annihilation cross-section is strongly peaked at $E_{\rm res} = m_V^2/2m_e$, typical positrons with $E_{e^+} > E_{\rm res}$ in the cascade must propagate through the material to lose energy through ionization and soft bremsstrahlung until they reach this resonance. 
During the propagation the $e^+$ undergoes MCS which modifies its direction; the direction of the positron \emph{at the interaction point} then determines the direction of the outgoing $V$, since the initial electron is nearly stationary and any initial state radiation is assumed to have negligible $p_T$. In previous work it was instead assumed that the initial positron direction determines the $V$ direction~\cite{Celentano:2020vtu}. 
Below we will show that the effect of MCS on the positron as it propagates between its birth and resonant annihilation into $V$ can have significant impact on experimental acceptances and must be treated carefully.

\begin{figure}[t]
    \centering
    \includegraphics[width=\linewidth]{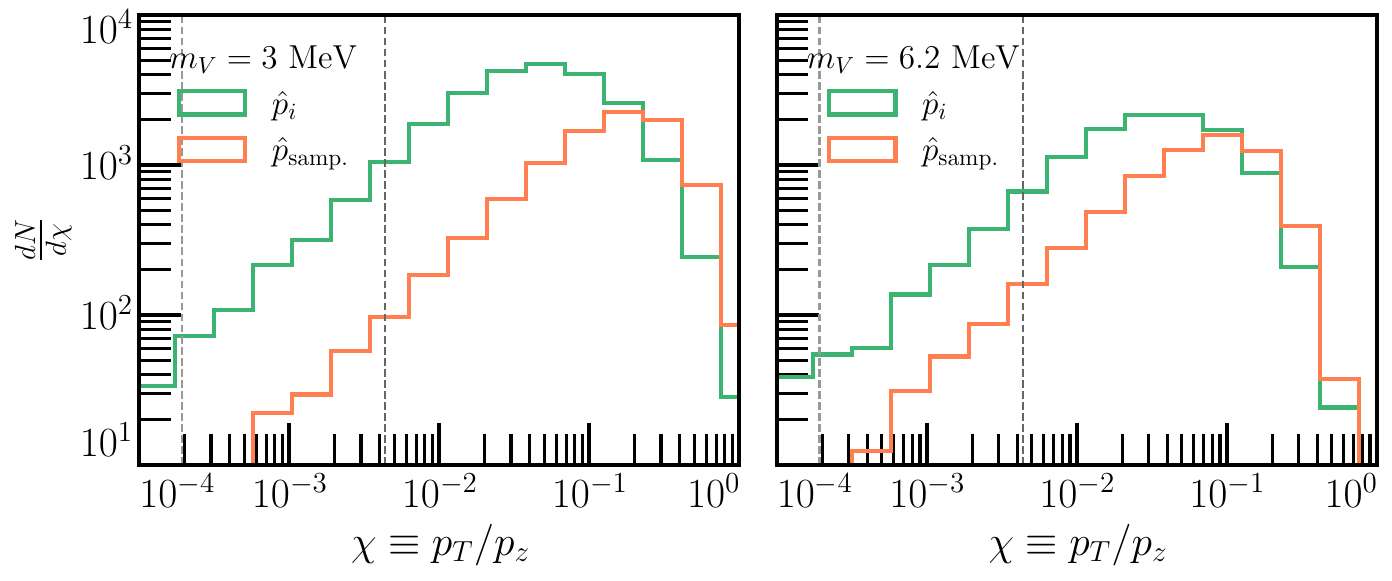}
    \caption{
    Comparison of $p_T/p_z$ (defined relative to the beam axis) for two different sampling procedures for positron annihilation for $m_V =3~{\rm MeV}$ (left) and $m_V=6.2~{\rm MeV}$ (right). 
    The green curve take the dark vector direction, $\hat{p}_V$, as the initial positron's direction, $\hat{p}_V=\hat{p}_i$, when it was created in the shower. 
    The orange curve implements the 
    direction of $V$ as the direction of the positron accounting for propagation and energy loss, $\hat{p}_V=\hat{p}_{\mathrm{samp}}$. 
    The left and right dashed vertical lines are representative of angular acceptances at ArgoNeuT (0.2 mrad) and DUNE (4.3 mrad), respectively.
\label{fig:Resonant_DirectionEffect} }
\end{figure}

In \cref{fig:Resonant_DirectionEffect}, we present the differential flux of resonantly produced dark vectors in a thick graphite target struck with 120 GeV protons for $m_V=3$~MeV (left) and 6.2~MeV (right). We simulate $\pi^0$ production from $pp$ collisions with \texttt{PYTHIA} and inject them into \PETITE, which decays $\pi^0\rightarrow \gamma \gamma$ and generates the EM shower.
The fluxes are given as a function of $p_T/p_z$, effectively the angular spread of the beam; vertical dashed lines represent the approximate angular acceptance of ArgoNeuT and a DUNE-like near detector, at 0.2 and 4.3 mrad. 
To illustrate the importance of propagating the positron down to the interaction point, we present two fluxes.
In green, we assume that the positron direction when it produces the dark vector through $e^+ e^- \to V(n\gamma)$ is given by its direction when it was created upstream in the shower, $\hat{p}_i$, leading to a dark vector direction $\hat{p}_{V} = \hat{p}_i$.
The orange curve demonstrates the impact of taking the $V$ direction as the direction of the positron when it annihilates into $V$ (i.e., it is sampled after propagation and energy loss), namely, $\hat{p}_V=\hat{p}_{\rm samp}$.
This propagation accounts for MCS which induces a non-negligible angular spread for the positron. 
We can clearly see that compared to the previous case $V$ tends to be created with significantly larger opening angles. 
In particular, properly accounting for MCS can reduce the  number of dark vectors within the geometrical acceptance of DUNE or ArgoNeuT by roughly an order of magnitude. Taking DUNE as a concrete example, the geometric acceptance differs by a factor of roughly $4$ for $m_V=6.2~{\rm MeV}$,  and by a factor of roughly $15$ for $m_V=3~{\rm MeV}$.

Besides properly modelling the angular spread, \PETITE also includes a careful treatment of singularities from soft and collinear photon emission by making use of QED parton distribution functions that resum QED corrections with leading-log accuracy. 
We do not attempt a ``matching and merging'' with fixed order results, leaving this for future work. 
The resulting cross section prediction differs from naive tree-level estimates by as much as ${\sim}20\%$, see \cref{A:radiative-return} for further details. 
Last, we note that we do not include atomic binding in our current simulations.

\subsection{Standard Model bremsstrahlung\label{subsec:sm_brem}}
Both bremsstrahlung and pair production are difficult processes to simulate due to their non-trivial phase space  singularity structures. 
The bremsstrahlung cross section is singular for collinear and soft photon emission, and in the low momentum transfer limit (see \cref{A:Landau,A:vegas} for more details).
Collinear singularities are regulated by the finite electron mass, while Coulomb singularities are regulated either by kinematic cuts or by atomic screening. The IR divergence due to soft photon emission is effectively cut-off by the minimum energy threshold of the shower. 
The difficulties of sampling from such a highly singular distribution, and the need for fast efficient code has lead the developers of popular tools for EM cascades such as \GEANTfour, \EGS~\cite{Hirayama:2005zm}, and \FLUKA~\cite{Ferrari:2005zk,Bohlen:2014buj}, to approximate kinematic samplings by default.\!\footnote{It should be noted that all three programs provide a more precise treatment of the {\it total} but not differential cross section than \PETITE (e.g., by including Coulomb corrections). 
\PETITE uses the tree-level Bethe-Heitler expressions for both the differential and total cross section.} 
We focus on bremsstrahlung although {\it default} implementations of pair production, which is related to bremsstrahlung by a crossing symmetry, have the same issues. 
In particular, in \GEANTfour, a full five-dimensional  sampling of phase space for $e^+e^-$ pair production is available using the Bethe-Heitler matrix element via the method \texttt{G4BetheHeitler5DModel} \cite{Bernard:2018hwf}.  

\begin{figure}[t]
    \centering 
    \includegraphics[width=\linewidth]{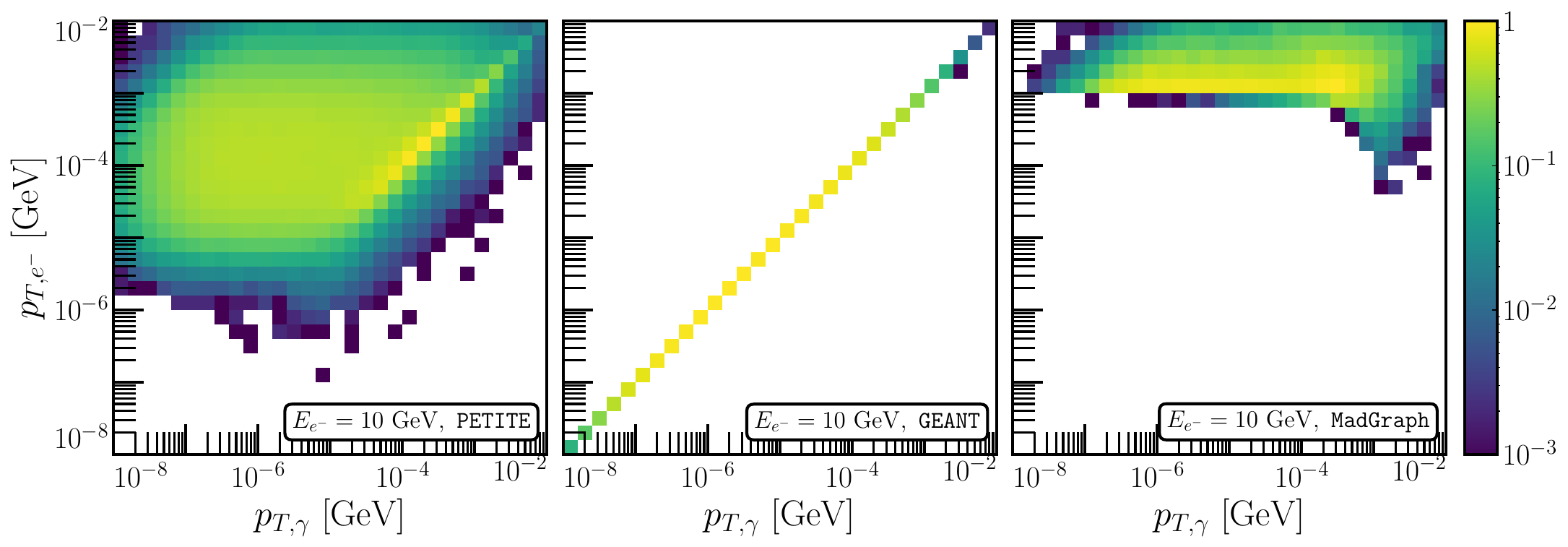}
    \caption{Transverse momentum distribution for daughter electrons and photons from bremsstrahlung produced by the first splitting of a $E_e=10~{\rm GeV}$ incident electron. 
    Distributions are computed with \PETITE (left), \GEANTfour (middle) and \madgraph with custom settings (right), see \cref{A:darkbremm}. 
    The implementation in \PETITE reproduces the analytic expressions in \cite{Berestetskii:1982qgu}. The artificially tight correlation between $p_T$ of the daughter $\gamma$ and daughter $e$ in \GEANTfour is readily seen, as is the lack of phase space coverage in \madgraph.  } \label{fig:brem-comp} 
\end{figure}

\EGS and \GEANTfour sample approximated distributions of photon energy and angle, and neglect momentum transfer to the nucleus, which leads to a spurious correlation between electron and photon transverse momentum (see Sec.~2.7.1 of \cite{Hirayama:2005zm} and Sec.~10.2 of \cite{GEANT_physicsmanual}).
\FLUKA uses the Berger-Seltzer parameterization which is two dimensional \cite{SELTZER198595,Ferrari:2005zk}. 
This reduction from a four-dimensional to two-dimensional phase space makes sampling much faster.  
The electron phase space, however, is then not properly correlated with the outgoing photon. To illustrate this point, we compare the correlation between electron and photon transverse momenta for the first splitting of a 10~GeV electron traversing graphite predicted by \PETITE, \GEANTfour and \madgraph in \cref{fig:brem-comp}.
In \GEANTfour, the final $p_T$ of the electron is set equal and opposite to the $p_T$ of the photon, essentially constraining the nuclear recoil to be vanishing (middle panel). 
In reality, these variables are not strongly correlated as can be seen in the left panel. \madgraph (right panel) lacks coverage of the low $p_T$ phase space, which is not a problem for predicting observables in high energy colliders such as the LHC, but in principle could affect dark sector production in EM showers.
We note that predicting events at such low $p_T$ region is not the \emph{raison d'\^{e}tre} of \madgraph, and in order to produce this panel, we have turned off all cuts in available in the run card.
Overall, bremsstrahlung is an unusually difficult process to simulate, due to its three-fold singularity structure. 

Although these differences are quite large for individual bremsstrahlung events, the effect on a bulk shower is much less pronounced due to angular spreading from MCS.
We have checked that the downstream effects of mismodelled bremsstrahlung $p_T$ distributions are not dramatic in thick targets but may be relevant for thin target experiments such as LDMX.

\subsection{Dark vs.~standard bremsstrahlung\label{subsec:sm_vs_ds_brem}}
\PETITE uses the exact tree-level dark bremsstrahlung differential cross-section to produce kinematics and to normalize rates. 
This allows us to highlight the differences of this process to its SM counterpart.
The matrix element for SM photon bremsstrahlung, $e^\pm Z\to e^\pm Z \gamma$, is strongly peaked towards small momentum transfers, leading to soft and collinear divergences.
The outgoing photons tend to be produced with small energy fraction and with small $p_T/p_z$ (the characteristic opening angles are $\theta_\gamma \sim m_e/E_e$, c.f.~\cref{eq:BremDSigma}).
For dark vector bremsstrahlung, on the other hand, nonzero $m_V$ regulates these divergences, leading to qualitatively different kinematics for $m_V> m_e$: the dark vectors tend to carry a significant fraction of the incoming $e^\pm$ energy and their emission angle is correlated with this fraction. 

\begin{figure}[t]
    \includegraphics[width=\linewidth]{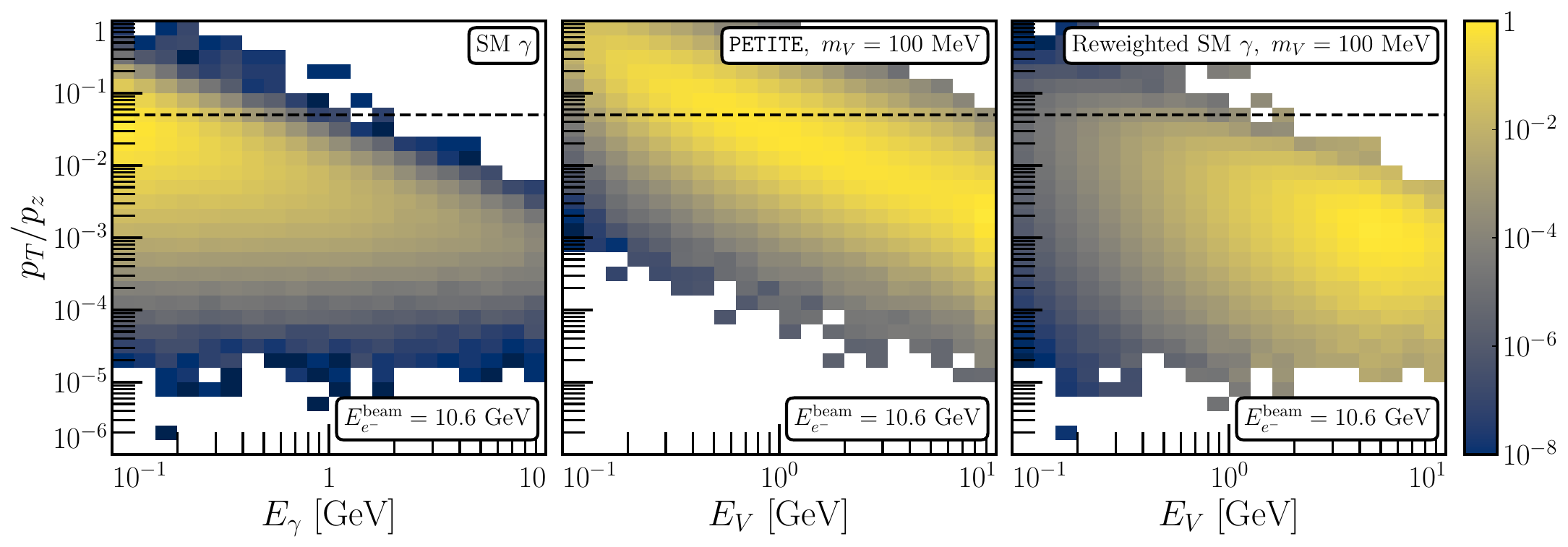}
    \caption{\label{fig:brem_pT_E}Comparison of reweighting vs resampling as a method for producing dark vector fluxes. The SM bremsstrahlung distribution from a 10.6 GeV $e^-$ is shown (\textbf{Left}), the dark vector distributions produced by reweighting the SM distribution using  \cref{reweight-ansatz} (\textbf{Centre}), and the distribution of dark vectors produced via \PETITE's resampling procedure (\textbf{Right}). A dashed line is drawn at $p_T/p_Z = 0.5/10$ to illustrate the BDX detector's angular acceptance. Each panel's colourscale is normalized to its distribution's highest-flux bin.} 
\end{figure}

While technical details can be found in \cref{A:Landau}, we illustrate this point in a concrete example. 
Let us take an experimental setup similar to the BDX experiment at Jefferson Lab~\cite{BDX:2016akw}.
We simulate an EM shower from a 10.6~GeV electron beam impinging on a thick target.
In the left panel of \cref{fig:brem_pT_E}, we show the kinematics ($p_T/p_z$ versus energy) of the photons produced via bremsstrahlung, as this is the dominant production mechanism.
The preference for soft and collinear photon emission can be clearly seen.
To compare to dark vector kinematics, we simulate the production of $m_V=100$~MeV dark photons in \PETITE.
For this mass, production due to hadronic reactions (e.g., $\pi^0\to\gamma V$),  positron annihilation and dark Compton scattering are all subdominant, leaving dark bremsstrahlung as the dominant channel.
In the middle panel of \cref{fig:brem_pT_E}, we present the kinematics for the dark vectors produced via dark bremsstrahlung. 
We can see two main differences with respect to photon kinematics: the $V$ spectrum is much harder, and there is a strong correlation between $E_V$ and its opening angle.
For reference, we plot a horizontal line corresponding to the BDX detector geometrical acceptance, a 50~cm~$\times$~50~cm target placed 10~m downstream from the back of the target~\cite{BDX:2016akw}.

Due to these kinematic differences, one cannot estimate the dark bremsstrahlung flux by simply re-weighting the photon bremsstrahlung flux in EM showers by the ratio of total cross sections.
To appreciate this, we present the dark bremsstrahlung flux obtained using a re-weighting procedure from Ref.~\cite{Capozzi:2021nmp} in the right panel of \cref{fig:brem_pT_E}.
This scheme uses the distribution of photons produced in a \GEANTfour simulation and applies a re-weighting factor bin-by-bin in energy and angle.  The number of photons, $N_\gamma(\theta_i,E_j)$,  produced through bremsstrahlung in the $(i,j)$-th bin centred around angle $\theta_i$ and energy $E_j$ is converted to the number of dark vectors, $N_V(\theta_i,E_j)$, in the same bin using the (angle-independent) ansatz 
\begin{equation}
    \label{reweight-ansatz}
    \begin{split}
    N_V(\theta_i,E_j) &= \kappaCoup^2 N_\gamma(\theta_i,E_j) \times f\qty(\tfrac{m_V}
    {E_e^{\rm est}(E_j)})~, \\
    f(x) &= {\rm min}\qty( 1 , 1154 \exp[ -24.32 x^{0.3174} ] )~,\\
    E_e^{\rm est}(E_\gamma)&=1.0773 E_\gamma + 13.716~{\rm MeV}~.
    \end{split}
\end{equation}
In the right panel of \cref{fig:brem_pT_E} we can see that the re-weighting procedure leads to mismodeling of the dark bremsstrahlung fluxes, with a much stronger preference for harder and more collinear dark vectors.
As we will see later, this can have significant impact on dark vector flux predictions in specific experimental setups.
The re-weighting procedure can overestimate the dark vector flux by 2 or more orders of magnitude at MiniBooNE or DUNE.
For example, for the BDX setup we consider in \cref{fig:brem_pT_E} the re-weighting procedure overestimates the dark vector geometric acceptance (i.e.~the fraction of the binned histogram below the dashed black line) by 35\% ($>99\%$ acceptance vs. 74\% acceptance). At the level of the overall rate we find that Ref.~\cite{Celentano:2020vtu} over-predicts the yield by roughly 5000 for $m_V=100~{\rm MeV}$ (we believe this comes from the normalization of $f(x)$ being off for large $m_V$).

\subsection{Impact on experimental yields \label{subsec:yields}}
\begin{figure}[t]
    \includegraphics[width=0.96\linewidth]{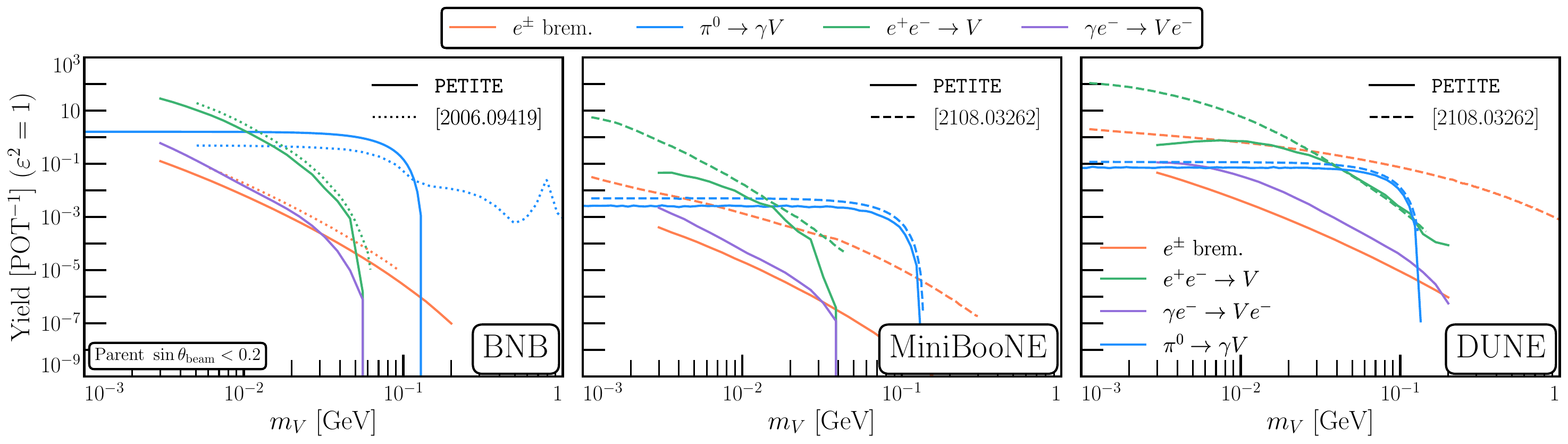}
    \caption{ Total yields of dark photons from electromagnetic secondary and hadronic primary production for three different experimental configurations. \textbf{(Left)} The flux emanating from the Booster Neutrino Beam with parent particles satisfying a cut of $\sin\theta_{\rm beam}<0.2$ on their angle relative to the beam axis. This configuration does not correspond to a specific detector and was chosen for direct comparison with Ref.~\cite{Celentano:2020vtu}. Flux of dark vectors passing through the MiniBooNE \textbf{(Centre)} and DUNE \textbf{(Right)} detectors, which have been chosen for direct comparison of \PETITE with Ref.~\cite{Capozzi:2021nmp}.  \label{fig:total-yield} } 
\end{figure}

In this subsection we compare existing results from the literature with the output from \PETITE, including all effects discussed above. 
We focus on the simplest figure of merit, which we take to be the total number of dark sector particles passing through a detector in experimental setups similar to MiniBooNE and DUNE.  In this subsection (and this subsection only) we consider a coupling of the dark vector $V_\mu$ to the hadronic electromagnetic current $J_\mu$ with the same coupling as electrons, i.e.~$|c_p|=|c_e|= c_\pi =\kappaCoup e$ in \cref{eq:interaction}.
This arises, for example, when considering kinetic mixing portals with dark photons.
We compare dark vector production in the Booster Neutrino Beam (BNB) for dark vector with parent particles' opening angles up to $\sin\theta_{\rm beam} < 0.2$ from \PETITE and Ref.~\cite{Celentano:2020vtu} in the left panel of \cref{fig:total-yield}.
The other two panels show \PETITE and Ref.~\cite{Capozzi:2021nmp} predictions for the dark vector flux at MiniBooNE (middle panel) and DUNE (right panel).

The output of \PETITE for dark bremsstrahlung agrees well with Ref.~\cite{Celentano:2020vtu}, as can be see in the left panel. 
Although our treatment of MCS in positron annihilation is different from Ref.~\cite{Celentano:2020vtu} its impact appears to be negligible for the large angular cut considered in that panel; 
for more realistic angular selection these differences are more apparent as we see in the middle and right panels.
The difference between the meson decay production curves (blue lines) in the centre and right panel are due to heavier mesons ($\eta$, $\eta'$) and proton bremsstrahlung, which we have not accounted for (we only considered $\pi^0$ decays). We have not been able to reproduce the direct hadronic production yield curve of Ref.~\cite{Celentano:2020vtu} as shown in the left panel  (\PETITE is roughly $2\times$ as large), but find that the photons which dominate the EM cascade are reasonably well modelled. We therefore find qualitative, but not perfect quantitative, agreement with Ref.~\cite{Celentano:2020vtu}.  
While secondary meson production affects the total rate of dark vector production for large opening angles, it is a subdominant effect when considering detectors far downstream from the target. 

The middle and right panels reveal discrepancies with Ref.~\cite{Capozzi:2021nmp}. 
As we discussed in \cref{subsec:sm_vs_ds_brem}, the re-weighting procedure for dark bremsstrahlung production (orange lines) used in Ref.~\cite{Capozzi:2021nmp} overestimates the dark vector production in the forward direction. 
By simulating the dark vector bremsstrahlung production using the appropriate matrix element, we find a much lower geometrical acceptance.
An immediate consequence of this is that dark bremsstrahlung is typically subdominant compared to other production mechanisms for masses between 1~MeV and 1~GeV in proton fixed target experiments.
Note that we still expect dark bremsstrahlung to be relevant in electron fixed target experiments for heavier masses where positron annihilation (green) is kinematically forbidden, particularly for hadrophobic models such as $L_e-L_\mu$ where dark bremsstrahlung can compete with primary production from meson decays. 

Another major finding of our study, is that positron annihilation rates (green lines) are modified substantially for detectors with small angular acceptances (see \cref{subsec:rad-return}). 
We reaffirm the findings of Refs.~\cite{Buonocore:2018xjk,Nardi:2018cxi,Celentano:2020vtu,Capozzi:2021nmp} that dark annihilation is an important production channel, and find reasonable quantitative agreement with their estimates for heavier dark vector masses. 
At low masses, we find discrepancies as large as a few orders of magnitude, which stem from the proper sampling of $p_T$ from MCS. 
We are able to reproduce the results of \cite{Capozzi:2021nmp} both using their methodology and a modified version of \PETITE that uses the initial positron momentum to determine the vector direction (c.f. \cref{subsec:rad-return}).  
Overall, due to small angular acceptances, we find that a proper simulation of dark bremsstrahlung and a proper sampling of dark vector $p_T$ in positron annihilation can change detector yields in realistic experiment setups by up to several orders of magnitude, depending on the mass of the dark vector.
This directly affects estimated experimental sensitivities to a wide range of models, including dark photons, $L_e-L_\mu$ and other flavour combinations, and $B-L$.
%

\section{Discussion \label{Conclusions}}

The prediction for dark sector fluxes emanating from a thick target is a highly non-trivial task due to the 
number and complexity of processes involved in realistic models. 
In this work we have tackled the problem of secondary production in an electromagnetic (EM) cascade. 
We have developed \PETITE ({\bf P}ackage for {\bf E}lectromagnetic {\bf T}ransitions {\bf I}n {\bf T}hick-target {\bf E}nvironments), a Python program which includes a lightweight EM shower generator, as well as a novel and universal algorithm for dressing (or resampling) generic SM Monte Carlo (MC) simulations with rare dark sector events. 
While we focused on estimating the production of dark vector particles in EM showers, our algorithm can be used for hadronic cascades, other BSM models or even for 
pure SM applications. In the latter case the resampling technique enables the simulation of rare but important events, 
providing a simple alternative to the MC biasing method~\cite{Ghinescu:2021sjm}. 
As resampling can be applied to any MC data (or implemented in any program), it can be easily interfaced with experimental simulation pipelines.

We have performed comparisons between \PETITE and other methods for predicting SM and BSM fluxes, extensively validating our approach and revealing several improvements enabled by \PETITE.
For example, \PETITE fully samples SM bremsstrahlung phase space, in contrast to existing tools such as \GEANTfour, \EGS, and \FLUKA which predict an incorrect correlation of the outgoing photon 
and electron momenta. The significance of this effect depends on the target thickness, as it can be washed out in thick targets. 
We also addressed outstanding disagreements in predictions of detector fluxes from BSM particle bremsstrahlung, finding that our first-principles-based simulation can correct 
potential signal yields by orders of magnitude.
Finally, we have identified a nontrivial interplay between the multiple Coulomb scatterings (MCS) and dark vector production via positron annihilation. Due to the angular spread induced by MCS and the small geometrical acceptance of realistic experimental setups, properly accounting for MCS in positron annihilation can change downstream yields by more than an order of magnitude.

Taking it all together, different qualitative conclusions can be gleaned from our work depending upon the primary incident particle on target.
Focusing on the specific scenario of a dark photon for concreteness, we conclude the following.
\begin{itemize}
    \item \textbf{Protons on target:} At lower dark photon masses, positron annihilation dominates the dark vector flux, but MCS broadening makes the flux lower than previously estimated.
    For higher masses, production is mostly from meson decays. 
    Dark bremsstrahlung rarely dominates the flux.
    \item \textbf{Electrons on target:}  Dark bremsstrahlung is important, and thus should be properly simulated accounting for the full phase space. 
    A substantial positron flux emerges after a few radiation lengths, and positron annihilation is relevant. As expected, hadronic production is negligible.   
\end{itemize}
We highlight that our findings are relevant for other dark sector models, such as $B-L$ or $L_e-L_\mu$.
In all cases, we have substantially improved theoretical control over the flux predictions of dark sector particles originating in EM showers. 
Other improvements can be implemented within \PETITE, such as the inclusion of the Landau-Pomeranchuk-Migdal effect relevant at high energies, more 
realistic models of ionization, and MCS, other BSM scenarios, and standard hadronic shower processes to extend our treatment to all showers.
Besides, \PETITE can also be useful to simulate rare standard model processes (e.g., $\mu^+\mu^-$ production) via the resampling technique developed here.

Looking forward, it will be important to perform fully realistic simulations of dark sector particle production in upcoming experiments, such as the SBN detectors, DUNE, DarkQuest, and BDX. 
Furthermore, although \PETITE offers a fast and lightweight solution to the generation of EM showers, it is meant to complement rather than replace more comprehensive generators such as \GEANTfour or \FLUKA. 
Therefore, it will be useful to interface the resampling algorithm presented herein with these tools, together with possible improvements on certain microphysics, e.g., SM bremsstrahlung.

\begin{acknowledgments}
We thank Peter LePage for help with the Python implementation of \vegas. We thank Adrian Thompson for helpful correspondences regarding the implementation of dark vector production in \cite{Capozzi:2021nmp}. We thank Bhaskar Dutta, Wes Ketchum, Shirley Li, and Stephan Meighen-Berger for useful discussions, and Brian Batell for feedback on the manuscript. RP is supported by the Neutrino Theory Network under Award Number DE-AC02-07CHI11359, the U.S. Department of Energy, Office of Science, Office of High Energy Physics under Award Number DE-SC0011632, and by the Walter Burke Institute for Theoretical Physics.
PJF and PANM are supported by Fermi Research Alliance, LLC under Contract No. DE-AC02-07CH11359 with the U.S.
Department of Energy, Office of Science, Office of High Energy Physics.
KJK is supported in part by DOE grant DE-SC0010813.
NB was supported in part by NSERC, Canada. We used \texttt{Package-X}~\cite{Patel:2015tea} and \texttt{FeynCalc}~\cite{Shtabovenko:2020gxv} to derive and check several analytic results in this work. Our numerical work was enabled by \vegas~\cite{Lepage:1977sw,Lepage:2020tgj}, \texttt{numpy}~\cite{harris2020array}, \texttt{scipy}~\cite{2020SciPy-NMeth} and~\texttt{matplotlib}~\cite{Hunter:2007}. 
\end{acknowledgments}

\appendix

\section{Review of QED processes \label{A:Landau}}
Processes involving two-to-two 
kinematics in QED are well known.  Here, for completeness, we supply the relevant differential cross sections for Compton scattering ($e^\pm \gamma \rightarrow e^\pm \gamma$), electron-positron annihilation ($e^-e^+\to\gamma\gamma$), Bhabha scattering ($e^+ e^- \rightarrow e^+ e^-$), and M\o ller scattering ($e^- e^- \rightarrow e^-e^-$):
\begin{align}
    \dv{t} \sigma_{\rm Comp.}&=\frac{8\pi \alpha^2 
    }{(s-m_e^2)^2}\qty[\qty(\frac{m_e^2}{s-m_e^2} + \frac{m_e^2}{u-m_e^2})^2 + \qty(\frac{m_e^2}{s-m_e^2} + \frac{m_e^2}{u-m_e^2})- \frac14\qty(\frac{s-m_e^2}{u-m_e^2} + \frac{u-m_e^2}{s-m_e^2})]~,\label{eq:compton} \\
    \dv{t} \sigma_{\rm Ann.}&= \frac{8\pi \alpha^2 
    }{s(s-4m_e^2)}\qty[\qty(\frac{m_e^2}{t-m_e^2} + \frac{m_e^2}{u-m_e^2})^2 + \qty(\frac{m_e^2}{t-m_e^2} + \frac{m_e^2}{u-m_e^2})- \frac14\qty(\frac{t-m_e^2}{u-m_e^2} + \frac{u-m_e^2}{t-m_e^2})]~,\\
    \dv{t}\sigma_{\rm Bhabha}&=\frac{4\pi \alpha^2 
    }{u(u-4m_e^2)}\qty[\frac{1}{t^2}\qty[\tfrac12(s^2+u^2)  + 4 m_e^2 (t-m_e^2)] + \frac{2}{tu}\qty(\tfrac12 s-m_e^2)(\tfrac12 s -3 m_e^2) + t\leftrightarrow u]~, \\
    \dv{t} \sigma_\text{M\o ller}&=  \frac{4\pi \alpha^2
    }{s(s-4m_e^2)}\qty[\frac{1}{t^2}\qty[\tfrac12(s^2+u^2)  + 4 m_e^2 (t-m_e^2)] + \frac{2}{tu}\qty(\tfrac12 s-m_e^2)(\tfrac12 s -3 m_e^2) + t\leftrightarrow u]~.\label{eq:moller}
\end{align}
We have taken the electrons, photons, and positrons to be unpolarized in all scattering processes. Note that $s,~t,$ and $u$ are defined according to the momentum ordering written for each process, $p_1 p_2 \to p_3 p_4$, where $s = (p_1 + p_2)^2 = (p_3 + p_4)^2$, $t = (p_3 - p_1)^2 = (p_4 - p_2)^2$, and $u = (p_3 - p_2)^2 = (p_4 - p_1)^2$.

Bremsstrahlung and pair-production in the field of a nucleus were first computed by Bethe and Heitler \cite{Bethe:1934za}; textbook discussions with modern notation can be found in Itzykson and Zuber \cite{Itzykson:1980rh}, and (in the greatest detail) in Landau and Lifshitz \cite{Berestetskii:1982qgu}. In the small-angle regime (which dominates) the differential cross section for bremsstrahlung process of $e^\pm Z \rightarrow e^\pm Z \gamma$ is given by \cite{Berestetskii:1982qgu}
\begin{equation}\label{eq:BremDSigma}
\begin{split}
    \frac{\dd \tilde{\sigma}}{\dd \tilde{\Pi}}= \frac{8}{\pi} \left\lvert F(\absQ^2)\right\rvert^2 \alpha^3 
    \frac{\epsilon' m_e^2}{\omega \epsilon \absQ^4} \delta\delta'\times \bigg\{ \frac{\delta^2}{(1+\delta^2)^2} + \frac{\delta'^2}{(1+\delta'^2)^2} &+ \frac{\omega^2}{2\epsilon \epsilon'} \frac{\delta^2+\delta'^2}{(1+\delta^2)(1+\delta'^2)} \\
    &\hspace{0.1\linewidth}- \qty(\frac{\epsilon'}{\epsilon} + \frac{\epsilon}{\epsilon'}) \frac{\delta \delta' \cos\phi}{(1+\delta^2)(1+\delta'^2)}\bigg\}~.
\end{split}
\end{equation}
Here the photon energy, in the lab frame, is $\omega$ and the incoming (outgoing) lepton energy, also in the lab frame, is  $\epsilon$ ($\epsilon^\prime= \epsilon-\omega$).  We cut off the IR range of the photon energy, but do not alter the upper bound from the maximum allowed kinematically, so $\omega_{\rm cut}<\omega<\epsilon-m_e$.
The variables $\delta$ ($\delta'$) are proxies for $\theta$ ($\theta'$) which are the angles between the outgoing photon and the incoming (outgoing) lepton.  They are defined as
\begin{equation}\label{eq:deltarelation}
         \delta^{(\prime)} = \frac{\epsilon^{(\prime)} \theta^{(\prime)}}{m_e}.
\end{equation}
The angles $\delta$ and $\delta'$ span $[0,\pi \times E_e^{(\prime)}/m_e]$; in the limit $m_e/E_e^{(\prime)}\rightarrow 0$ this tends to the positive real axis $[0,\infty)$.
The azimuthal angle between the incoming and outgoing lepton is denoted $\phi$ and lies in the range $0\le\phi<2\pi$.
The momentum transfer to the nucleus, $Q^2=-q_\mu q^\mu = \vb{q}^2$, can be expressed as
\begin{equation}\label{eq:BremQSq}
    \absQ^2 = m_e^2\left[\left(\delta^2 + \delta'^2 - 2\delta\delta' \cos\phi\right) + m_e^2 \left( \frac{1+\delta^2}{2\epsilon} + \frac{1+\delta'^2}{2\epsilon'}\right)^2\right].
\end{equation}
The total phase space $\dd \tilde{\Pi} = \dd \omega \dd \delta \dd \delta' \dd \phi$. 
We also explicitly include an atomic form factor $\left\lvert F(\absQ^2)\right\rvert^2$~\cite{Schiff:1951zza,Tsai:1973py} which screens interactions at small $\absQ^2$ -- it has the behaviour that for small $\absQ^2$, $\left\lvert F(\absQ^2)\right\rvert^2 \propto \absQ^4$ (eliminating the $\absQ^{-4}$ singularity present in \cref{eq:BremDSigma}) and for large $\absQ^2$, it scales like $Z^2$ (the unscreened case). The effect of screening is most apparent for large-$Z$ targets and large incident $\epsilon$.
See~\cref{A:formfactors} for explicit definitions of these form factors and the appropriate simplifications we take for SM processes.

Pair-production is related to bremsstrahlung by crossing symmetry and therefore it may be described using a similar set of approximations. It is the dominant process for sufficiently high-energy photons propagating through a thick target~\cite{ParticleDataGroup:2022pth}. 
The differential cross-section in the small-angle approximation is given by~\cite{Berestetskii:1982qgu}
\begin{equation}\label{eq:PairProdDSigma}
\begin{split}
    \frac{\dd \tilde{\sigma}}{\dd \tilde{\Pi}}= \frac{8}{\pi} \left\lvert F(\absQ^2)\right\rvert^2 \alpha r_e^2 \frac{m_e^4 \epsilon_+ \epsilon_-}{\omega^3 \absQ^4} \delta_+ \delta_-\times \bigg\{ -\frac{\delta_+^2}{(1+\delta_+^2)^2} - \frac{\delta_-^2}{(1+\delta_-^2)^2} &+ \frac{\omega^2}{2\epsilon_+ \epsilon_-} \frac{\delta_+^2+\delta_-^2}{(1+\delta_+^2)(1+\delta_-^2)} \\
    &+ \qty(\frac{\epsilon_+}{\epsilon_-} + \frac{\epsilon_-}{\epsilon_+}) \frac{\delta_+ \delta_- \cos\phi}{(1+\delta_+^2)(1+\delta_-^2)}\bigg\}.
\end{split}
\end{equation}
Here, the incoming photon has energy $\omega$, and the outgoing electron (positron) has energy $\epsilon_-$ ($\epsilon_+ = \omega - \epsilon_-$). Similar to \cref{eq:BremQSq}, $q$ is the momentum transferred to the nucleus and in this approximation, the expression for $\absQ^2$ is
\begin{equation}
    \absQ^2 = m_e^2\left[ \left(\delta_+^2 + \delta_-^2 + 2\delta_+ \delta_- \cos\phi\right) + m_e^2\left(\frac{1+\delta_+^2}{2\epsilon_+} + \frac{1+\delta_-^2}{2\epsilon_-}\right)\right].
    \label{eq:PairProdQSq}
\end{equation}
The integration phase space is $\dd \tilde{\Pi} = \dd \epsilon_+ \dd \delta_+ \dd \delta_- \dd \phi$. The energy $\epsilon_+$ spans the range $[m_e, \omega - m_e]$. The other three variables have the same range as for electron bremsstrahlung, and $\delta_{\pm}$ are related to the outgoing positron/electron angles $\theta_{\pm}$ through a similar relation as \cref{eq:deltarelation}. The atomic form factor $\left\lvert F(\absQ^2)\right\rvert^2$ is the same as the above, as well. For both pair production and bremsstrahlung a nuclear form factor can be similarly included, but its effect is minimal because of the dominance of small $\absQ^2$ where $F_{\rm nucl.} (\absQ^2) \approx 1$. This is not necessarily true for dark photon production, which we treat separately in \cref{A:darkbremm}. 

For high-$Z$ nuclei, Coulomb corrections modify the cross section \cite{Bethe:1954zz,Davies:1954zz,Tsai:1973py}. We have not included this effect within \PETITE, which means that the \PETITE mean free paths are somewhat shorter than in nature. The advantage of not including Coulomb corrections is that the theory is entirely self-consistent (i.e., the total and differential cross sections derive from the same formulae).   

\section{Form factors}\label{A:formfactors}

Depending on the momentum exchanged when a particle interacts with material, the scattering may be coherent off the entire atom, nucleus, a part thereof, or off an atomic electron.  The form factor 
$F(\absQ^2)$, accounts for both atomic screening at low $\absQ^2$ and the loss of nuclear coherence at large $\absQ^2$.
It contains both a coherent (elastic) contribution \cite{Tsai:1973py, PhysRevD.8.3109}, which scales as the square of the nuclear charge, and an incoherent (inelastic) contribution which scales linearly with $Z$ 
\begin{equation}
    |F(\absQ^2)|^2 = Z^2 G_{\rm el}(\absQ^2) + Z G_{\rm inel} (\absQ^2)~.
    \label{F_def}
\end{equation}
Where the two contributions are~\cite{Bjorken:2009mm}
\begin{align}
    G_{\rm el}(\absQ^2)  &= \qty(\frac{a^2 \absQ^2}{1+ a^2\absQ^2})^2\qty(\frac{1}{1+\absQ^2/d})^2~, \label{G_el}\\
    G_{\rm inel} (\absQ^2) &= \qty(\frac{a^{\prime2} \absQ^2}{1+ a^{\prime2}\absQ^2})^2 \qty(\frac{1+ \frac{\absQ^2}{4 m_p^2}(\mu_p^2-1)}{(1+ \absQ^2/\Lambda_H^2)^4})~.
    \label{G_inel}
\end{align}
In each case the form factor is a product of terms accounting for atomic and hadronic effects. For $G_{\rm el}$ these factors encode atomic screening of nuclear charge and the finite size of nucleus, while for $G_{\rm inel}$ they describe atomic excitation and finite proton size.
The atomic parameters are given by 
\begin{equation}
    a=\frac{111}{Z^{1/3}m_e}  \qq{and}   a'= \frac{773}{Z^{2/3}m_e} ~,
    \label{eq:atomic_para}
\end{equation}
while the hadronic parameters are
\begin{equation}
\Lambda_H=0.84~{\rm GeV}  \qq{and} d=\left(0.4 A^{-1/3}\, \mathrm{GeV} \right)^2~.
\end{equation}
The proton's magnetic moment is $\mu_p=2.79$, and $m_p$ is the mass of the proton. 
Notice that the second bracketed term in \cref{G_inel} is {\it not} squared,
see Footnote 4 of Ref.~\cite{Celentano:2020vtu}.

Atomic screening is important for SM bremsstrahlung and pair production. When considering pair production of electrons, one almost never encounters $\absQ^2$ large enough that the nuclear form factor, or incoherent scattering contributes. Therefore, in \PETITE for SM interactions, $G_{\rm inel}$ is set to zero, and only the atomic form factor in included in $G_{\rm el}(\absQ^2)$ (i.e.\ $d\to\infty$). Appropriate modifications to include the loss of nuclear coherence and incoherent scattering could be easily added if so desired (this would be important, for instance, if one wanted to consider $\gamma Z \rightarrow \mu^+\mu^- Z$).

\section{Continuous processes\label{A:continous_processes}}
\subsection{Energy loss \label{A:energy-loss}}
In \PETITE we have not attempted a stochastic treatment of energy loss.\footnote{In principle such a description would involve proper sampling from the Landau-Vavilov distribution. This is computationally expensive and is therefore not implemented in standard codes such as \GEANTfour, which instead employ toy models of ionization energy loss.} We instead treat energy loss with a deterministic function $\langle \dd E/\dd x\rangle$ which includes contributions from soft bremsstrahlung (photon emissions below the cut $E_{\rm min}$ -- see \cref{subsec:implementation_of_petite}) and ionization. Because we take $E_{\rm min}\sim \MeV$, well below the critical energy for most materials~$\mathcal{O}(10-100\;\MeV)$~\cite{ParticleDataGroup:2022pth},  $\langle \dd E/\dd x\rangle$ is approximately an energy-independent constant since hard bremsstrahlung (photon emissions above $E_{\rm min}$) is explicitly simulated as a discrete process.

The propagation of a particle between its birth and hard interaction takes place in a series of 
steps, with each step much smaller than the mean free path. After each step the energy is adjusted using the average energy loss; if the energy falls below $E_{\rm min}$, the particle is terminated and will not be propagated anymore.

\subsection{Multiple Coulomb scattering \label{A:multiple-scatter}}
MCS is crucial for the proper modelling of shower dynamics, especially its transverse distribution. \PETITE has two different implementations of MCS available, and the ability to tune the typical scattering angle from MCS.  In what follows all formulae are written for particles with unit charge. In both the Lynch-Dahl and Bethe-Moli\`ere formalisms \cite{Bethe:1953va,Lynch:1990sq}, rescaling the outputted MCS angle can be interpreted as re-scaling the characteristic scattering angle $\chi_c$. 
\paragraph{Bethe-Moli\`ere}
The slower implementation in \PETITE is based on the theory of multiple scattering first developed by Moli\`ere \cite{Moliere:1947zza}, and then simplified and improved by Bethe \cite{Bethe:1953va}. In practice, MCS in this formalism is implemented using an asymptotic expansion in a parameter $B$, defined by the transcendental equation 
\begin{equation}
    B+\ln B= \ln \frac{\Omega}{1.167}~, 
\end{equation}
where 
\begin{equation}
    \Omega = \frac{\chi_c^2}{\chi_a^2} = \frac{7800~ t ~(Z+1) Z^{1/3} }{\beta^2 A \qty[1+ 3.34 \qty(\frac{Z \alpha}{\beta})^2 ]}
    \label{eq:lynch_dahl_omega_def}
\end{equation}
with the characteristic scattering, $\chi_c$, and screening, $\chi_a$, angles being defined below \cref{eq:bethe_transcendental}. Here $t$ is the thickness of material traversed in g/cm$^2$, $\beta$ the particle velocity, $Z$ the atomic number and $A$ the atomic weight in g/mol. $\Omega$ is the mean number of scatters, so typically $\Omega\gg 1$ unless $t < 10^{-3} X_0$~\cite{Lynch:1990sq}.

The Moli\`ere distribution is given by 
\begin{align}
    f(x) &= f_0(x) + B^{-1} f_1(x) + \ldots \\ 
    f_0(x)& = \e^{-x}~, \\
    f_1(x)&= \e^{-x}(1-x)\qty[\log(x)-{\rm Ei}(x) + (1-2\e^{-x})]~.
\end{align}
The probability distribution function for scattering through an angle $\theta$ is then
\begin{equation}
    P(\theta)~ \theta \dd \theta  =   f\qty(\vartheta^2)~\vartheta \dd \vartheta, 
\end{equation}
where $ \vartheta=\theta/\sqrt{B\chi_c^2}$. 
Note that $f_i$ for $i>0$ correspond to 
non-Gaussian corrections that come from rare, relatively large-angle deflections. Because 
of this non-Gaussianity, numerical sampling of Bethe-Moli\`ere distribution is relatively slow. 
\paragraph{Lynch-Dahl (Gaussian)}
In many applications a Gaussian approximation to MCS can be used, enabling very fast sampling. The default settings in \PETITE use the Lynch-Dahl parametrization where the angles are sampled from a normal distribution with variance  \cite{Lynch:1990sq}, 
\begin{align}
    \sigma^2&= \frac{\chi_c^2}{1+F}\qty[ \frac{1+v}{v} \log(1+v)-1]~,\\
    v &= \frac{\Omega}{2(1-F)}~. 
\end{align}
where $\Omega$ is given in \cref{eq:lynch_dahl_omega_def} and we take $F=0.98$ in \PETITE.

\section{Sampling and \vegas \label{A:vegas} }
For the sake of integration and event sampling of SM bremsstrahlung and pair production, we perform a change of coordinates relative to $\dd \tilde{\Pi}$ (defined below \cref{eq:BremQSq} and \cref{eq:PairProdQSq}). For bremsstrahlung, the new coordinates are defined in \cref{tab:BremIntegrationXs}. Since the integrand in \cref{eq:BremDSigma} is singular for small $q^2$ and scales like $1/\omega$, it peaks when $\delta \approx \delta'$ and when $\delta,\delta', \omega \rightarrow 0$. Also, by converting to dimensionless integration variables, we are more readily able to recycle trained \texttt{VEGAS} integration maps for new incident electron/positron energies.
\begin{table}[h]
    \centering
    \caption{Integration variables $x_i$ used for training \texttt{VEGAS} on SM bremsstrahlung emission for electrons and positrons. The right column provides the mapping between these and the kinematics variables used in \cref{eq:BremDSigma}.}
    \label{tab:BremIntegrationXs}
    \begin{tabular}{c||c}\hline
        \texttt{VEGAS} Integration Variable -- SM Bremsstrahlung &  Mapping to \cref{eq:BremDSigma}\\ \hline\hline
         $x_1 \in [0,1]$ & $\omega = E_{\gamma}^{\rm min.} + x_1\left(\epsilon - m_e - E_{\gamma}^{\rm min.}\right)$ \\ \hline
         $x_2 \in [0,2]$ & $\delta = \frac{\epsilon}{2m_e} \left(x_2 + x_3\right)$ \\ 
         $x_3 \in [-2,2]$ & $\delta' = \frac{\epsilon}{2m_e} \left(x_2 - x_3\right)$ \\ \hline
         $x_4 \in [0,1]$ & $\phi = 2\pi \left(x_4 - \frac{1}{2}\right)$ \\ \hline
    \end{tabular}
\end{table}

We also perform a parameter redefinition for pair production to improve the efficiency and accuracy of event sampling, with the mapping provided in \cref{tab:PairProdDSigma}. This has the same advantages as in the bremsstrahlung case due to the $q^{-4}$ nature of \cref{eq:PairProdDSigma}.

Finally, the inclusion of atomic screening tames the low-$q^2$ Coulomb singularity. This renders pair-production and bremsstrahlung numerically stable. The inclusion of atomic screening is also physical, and modifies the cross section and average angles of emission for both processes.

\begin{table}[h]
    \centering
    \caption{Integration variables $x_i$ used for training \texttt{VEGAS} on SM pair production. The right column provides the mapping between these and the kinematics variables used in \cref{eq:BremDSigma}.}
    \label{tab:PairProdDSigma}
    \begin{tabular}{c||c}\hline
        \texttt{VEGAS} Integration Variable -- SM Pair Production &  Mapping to \cref{eq:PairProdDSigma}\\ \hline\hline
         $x_1 \in [0,1]$ & $\epsilon_+ = m_e + x_1\left(\omega - 2m_e\right)$ \\ \hline
         $x_2 \in [0,2]$ & $\delta_+ = \frac{\omega}{2m_e} \left(x_2 + x_3\right)$ \\ 
         $x_3 \in [-2,2]$ & $\delta_- = \frac{\omega}{2m_e} \left(x_2 - x_3\right)$ \\ \hline
         $x_4 \in [0,1]$ & $\phi = 2\pi x_4$ \\ \hline
    \end{tabular}
\end{table}

\section{Validation of SM showers in \PETITE \label{A:SM_shower_validation}}
In this section we compare SM showers generated with \PETITE to predictions from analytic theory~\cite{Tsai:1966js} and from \GEANTfour.

\subsection{Comparison with analytic shower theory}
The analytic shower theory developed by Tsai and Whitis in Ref.~\cite{Tsai:1966js} included only bremsstrahlung and pair production with approximate differential rates in order to obtain tractable integro-differential equations for particle intensities as a function of depth and energy. These equations were solved iteratively for sequential generations of particles starting from an initial electron beam. 

\begin{figure}
    \centering
    \includegraphics[width=0.9\textwidth]{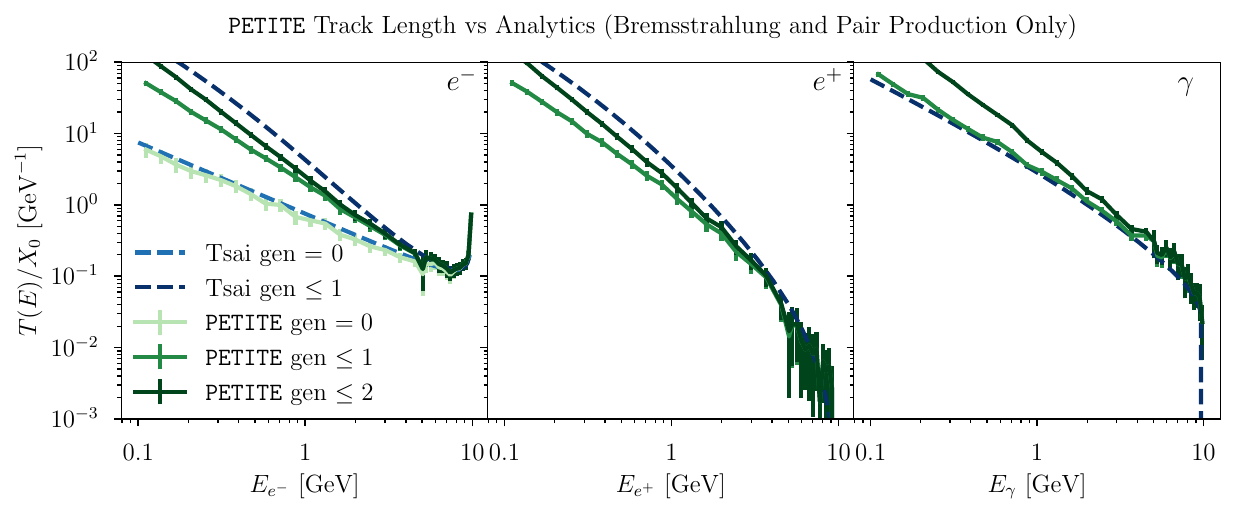}
    \caption{Track length distributions for $e^-$, $e^+$ and $\gamma$ for an initial $10\;\GeV$ $e^-$ beam on a thick graphite target. Each panel compares predictions of \PETITE to the analytic calculations of Tsai and Whitis~\cite{Tsai:1966js}. All processes except for bremsstrahlung and pair production have been turned off in \PETITE to closely match the assumptions of Ref.~\cite{Tsai:1966js}. The \PETITE and analytic theory curves are labelled by the generation number, which counts the number of parent photons a particle has had: generation 0 corresponds to the primary beam that lost energy due to bremsstrahlung, generation 1 includes photons produced from the primary beam and $e^\pm$ they generate. The analytic theory describes well the straggling of the primary beam and first generation photons, but it systematically over-predicts $e^\pm$ pair production at lower energies. The bars on the \PETITE results are statistical uncertainties for $10^3$ showers.}
    \label{fig:petite_vs_tsai_brem_and_pair_prod_only}
\end{figure}

In \cref{fig:petite_vs_tsai_brem_and_pair_prod_only} we turn off all processes except for bremsstrahlung and pair production in \PETITE and compare track lengths to analytic theory generation by generation, for an initial $10\;\GeV$ electron on a graphite target. The track length for particle type $i = \{e^-, e^+, \gamma\}$ is  
\begin{equation}
T_i(E) = X_0\int_0^T \dd t I_i(t, E),
\end{equation}
where $t$ is the depth inside the target in radiation lengths, $I_i(t, E)$ is the differential particle intensity, i.e. $\int dE I_i(t,E)$ is the total current passing through a surface at depth $t$.
The track length is a useful quantity in practice because the total yield of dark sector particles is proportional to $\int \dd E T_i(E) \sigma(E)$. We can therefore think of $\dd E T_i(E)$ as the cumulative distance travelled by all particles of type $i$ with energy in the interval $[E,E+\dd E]$ in the entire cascade (i.e., it takes into account not only the actual path length of individual particles, but also their multiplicity in the shower). The track length is extracted numerically from \PETITE following the prescription from Ref.~\cite{Marsicano:2018krp}:
\begin{equation}
    T_i(E) = X_0 \Delta t\sum_j I_i (t_j, E),
\end{equation}
where $I_i(t_j,E)$ is the ``measured'' intensity at depth $t_j$ (the depths at which the intensities are calculated are spaced by $\Delta t$). We used $10^4$ equally-spaced $t_j$'s but the numerical results are stable to variations of this number by a factor of ${\sim}10$.

The main features of \cref{fig:petite_vs_tsai_brem_and_pair_prod_only} are that the analytic theory provides an excellent description of 0th generation electrons (the beam electrons that have undergone any number of bremsstrahlung events) and 1st generation photons (the photons produced in those same bremsstrahlung events). However, it significantly over-predicts first generation electrons and positrons ($e^\pm$ that are pair produced from 1st generation photons). We have tracked this issue to the ``complete screening'' approximation employed in Ref.~\cite{Tsai:1966js}. This approximation replaces the energy-dependence of the cross-section $\sigma(E)$, by its asymptotic value at large energies, $\sigma_{\rm asym}$, which is absorbed into the definition of the radiation length. Tsai and Whitis scale all their length scales to radiation lengths, whereas in \PETITE each step is determined using the full energy-dependent mean free path $\lambda(E)$. Since $\sigma(E)\leq \sigma_{\rm asym}$ we have $\lambda(E)\geq \lambda_{\rm asym}$. This results in a suppression of the normalized track length distribution $T(E)/X_0$ that is energy dependent. The over-prediction of secondary $e^\pm$ pairs compared to simulations is also evident in previous work, see, e.g., Fig. 4 of Ref.~\cite{Marsicano:2018krp}.

\begin{figure}
    \centering
    \includegraphics[width=0.9\textwidth]{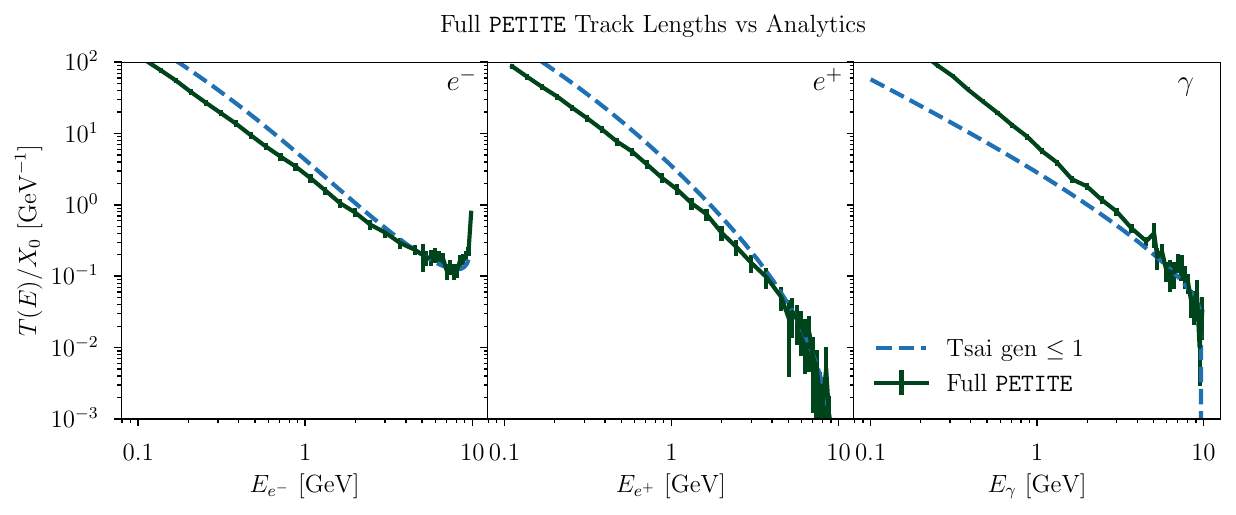}
    \caption{Same as \cref{fig:petite_vs_tsai_brem_and_pair_prod_only}, but with all processes enabled in \PETITE. }
    \label{fig:full_petite_vs_tsai}
\end{figure}

It is also interesting to consider the impact of all other processes that are present in \PETITE, but not in Ref.~\cite{Tsai:1966js}: these include annihilations, Compton, M\o ller and Bhabha scattering, ionization energy loss and all photon emissions (without generation restrictions). In \cref{fig:full_petite_vs_tsai} we compare \PETITE with  all of these processes (``full \PETITE'') enabled to the analytic theory. We see that the additional physics does not qualitatively impact the conclusions of previous paragraph; the main difference is that because \PETITE computes all $\gamma$ emissions, the photon track length is larger than that predicted by Ref.~\cite{Tsai:1966js} (this is already evident in the right panel of \cref{fig:petite_vs_tsai_brem_and_pair_prod_only}).

\begin{figure}
    \centering
    \includegraphics[width=0.9\textwidth]{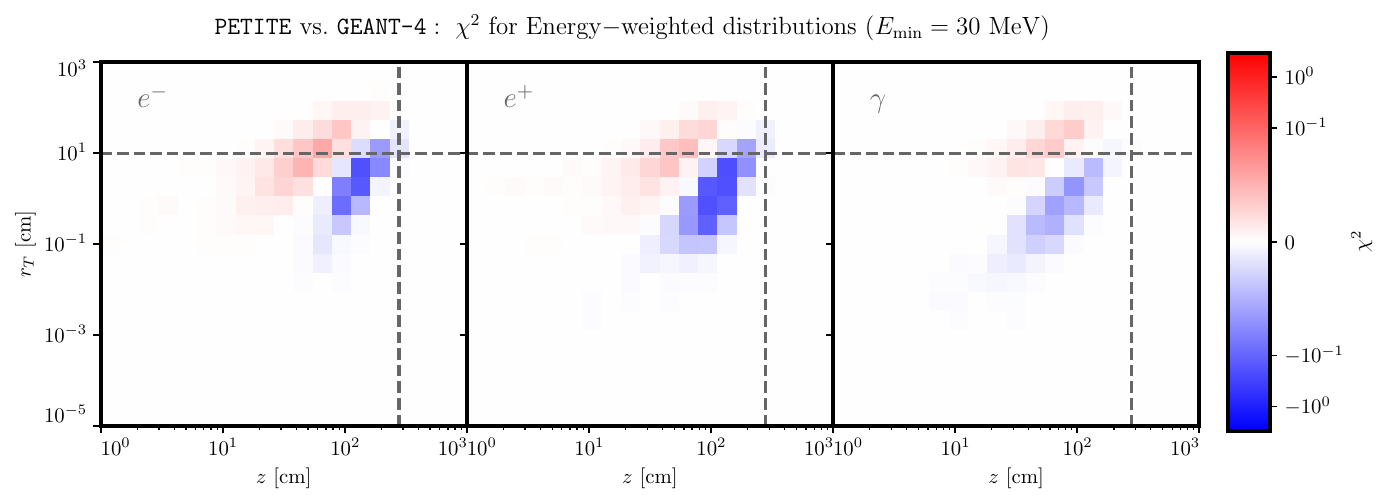}\\
    \includegraphics[width=0.9\textwidth]{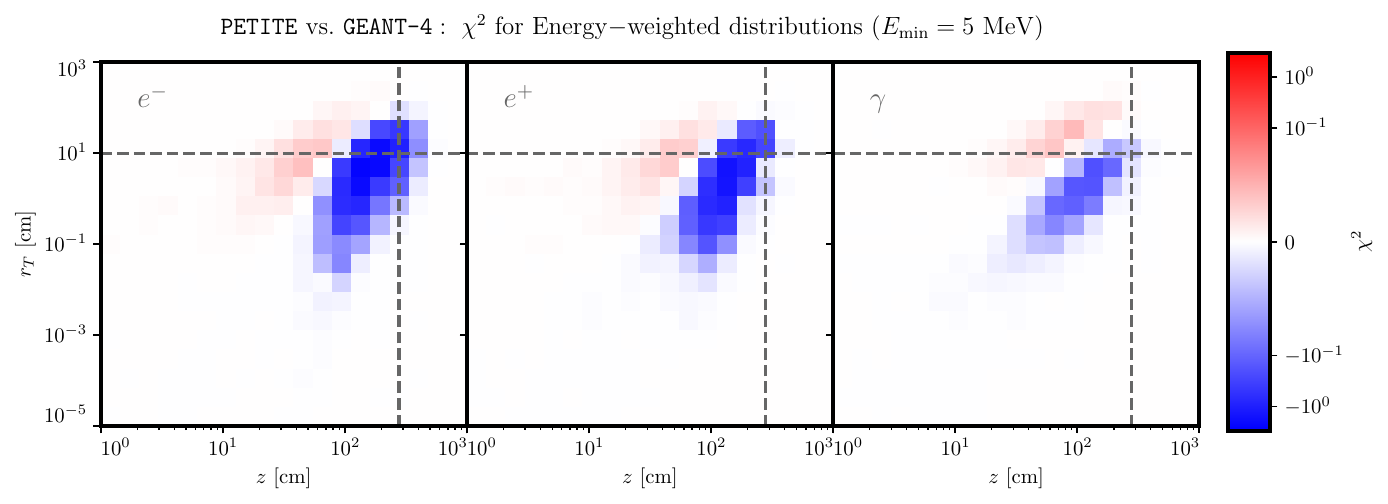}  
    \caption{Signed $\chi^2$ test statistic for \PETITE and \GEANTfour predictions of $dE/dr_T d z$, the energy-weighted single-shower particle distribution in transverse ($r_T$) and longitudinal distance ($z$) from the 
    beam axis, for a 10 GeV photon impinging on a graphite target.  Each column corresponds to $e^-$, $e^+$ and $\gamma$ respectively. The two rows use different values of $E_{\mathrm{min}}$, the minimum particle energy tracked in the shower. The $\chi^2$ values of $\mathcal{O}(1)$ indicate similarity of distributions (i.e., within statistical uncertainties).  The vertical and horizontal dashed lines indicate estimated $95\%$ shower energy containment in the longitudinal and transverse directions; they correspond to $\sim 14.4X_0$ and $2 r_{\mathrm{Moliere}}$, respectively~\cite{Fabjan:2003aq}. 
    }
    \label{fig:petite_vs_geant_chi2}
\end{figure}
\subsection{Comparison with \GEANTfour}
In many experiments, dark sector signal yield is maximized when the parent SM particle has a small transverse momentum and a large energy. Thus, modelling this part of the SM cascade is crucial. In \cref{fig:petite_vs_geant_chi2} we show a comparison of energy-weighted shower distributions (in transverse and longitudinal distances $r_T$ and $z$) between \PETITE and \GEANTfour, taking an incoming beam of $10$ GeV photons in a graphite target. The statistical significance of differences between the two distributions is estimated by computing a signed $\chi^2$ for average particle energies ($\overline{E}$) in each $r_T$ and $z$ bin:
\begin{equation}
    \chi^2(r_T, z) = \sgn\left(\overline{E}(\PETITE)-\overline{E}(\GEANTfour)\right)\frac{(\overline{E}(\PETITE)-\overline{E}(\GEANTfour))^2}{\sigma_{E}^2(\PETITE)+\sigma^2_{E}(\GEANTfour)} 
\end{equation}
where $\sigma_{E}^2$ is the variance of $\overline{E}$. The particle number in each bin is sensitive to the minimum tracking energy $E_{\mathrm{min}}$; smaller tracking energies imply more low-energy particles and greater sensitivity to low-energy physics. Therefore in \cref{fig:petite_vs_geant_chi2} we show two different choices of $E_{\mathrm{min}}$. The trend is that earlier parts of the showers (smaller $r_T$ and $z$ which are usually the most relevant for dark sector production) are consistent between the two programs since generally $\chi^2 \ll 1 $ in each bin. Statistically-significant differences emerge at large transverse and longitudinal distances when we lower $E_{\mathrm{min}}$. We have identified the likely culprit of these differences to be modelling of multiple Coulomb Scattering; \GEANTfour implements separate MCS models for low- and high-energy particles, and it always includes rare high-deflection events, whereas \PETITE uses a single model for MCS. We have checked, for example, that turning off MCS completely in both simulations (an unphysical limit) or using the Bethe model in \PETITE (which includes the rare high-angle scatterings but is noticeably slower than a simple Gaussian model) improves agreement.   
We emphasize that for most experimental set-ups that feature a distant, on-axis detector these differences are immaterial, since the signal events will be generated by the earlier, energetic and forward-going part of the cascade. The \PETITE and \GEANTfour distributions in this region of phase space are in very good agreement.

\section{Dark vector bremsstrahlung \label{A:darkbremm}}
The mass of the dark sector particles qualitatively changes the energy and angular distribution of its emission compared to its SM counterpart. These important features are determined by the soft and collinear singularity structure of the amplitude. This means that accurate simulation of this interaction can be numerically challenging despite the simplicity of the underlying process. In this Appendix we validate our implementation of dark bremsstrahlung and compare it to existing tools.

Neglecting mass-suppressed diagrams where the dark vector couples to the nucleons, the tree-level amplitude for $e(p) + N (P) \to e(p') + N(P') + V(k)$ is~\cite{Liu:2017htz}
\begin{equation}
\mathcal{M} = e^2 \kappaCoup \frac{F(\absQ^2)}{-\absQ^2}\varepsilon_\mu(k)^*\bar{u}(p') \left[(\slashed{P}+\slashed{P}') \frac{\iu (\slashed{p}-\slashed{k} + m_e)}{m_V^2 - 2 p \cdot k }\gamma^\mu +  \gamma^\mu\frac{\iu (\slashed{p}'+\slashed{k} + m_e)}{m_V^2 + 2 p' \cdot k }(\slashed{P}+\slashed{P}')\right]u(p)~.
\end{equation}
where $q=P'-P$ and $\absQ^2=-q^2$. Spin-summing and averaging the squared the matrix element, one obtains an integral expression for the differential cross section \cite{Tsai:1989vw,Gninenko:2017yus}
\begin{equation}
    \frac{\dd\sigma}{\dd E_{V}\dd\cos\theta} = 
    \left(\frac{\kappaCoup^2 \alpha^3|\vec{k}|}{8M_N^2|\vec{p}||\vec{k}-\vec{p}|}\right)\int_{Q^2_{\rm min}}^{Q^2_{\rm max}}\frac{\dd \absQ^2}{Q^4}|F(Q^2)|^2I_{\phi}(t, E_{V}, \cos\theta; M_N, m_{V}, E_e)~.
\end{equation}
Here $I_{\phi}$ is the result of integrating the matrix element squared over the azimuthal angle of the exchanged momentum $\vec{q}$  and $E_0$ is the initial electron's energy in the nucleus rest frame.  The full expressions for $I_\phi, Q_{\rm min,max}$ are presented in Ref.~\cite{Tsai:1989vw,Gninenko:2017yus}. 

The dark sector and beam particle particle masses,  $m_V$ and $m_e$, regulate the soft and collinear divergences, which can be seen explicitly in the kinematic regime with $q^2 \approx -\absQ^2_\mathrm{min}$~\cite{Bjorken:2009mm}, where the minimum photon virtuality is
\begin{equation}
Q_\mathrm{min} \approx \frac{m_V^2}{2E} = 0.5\;\mathrm{keV}\;\;\bigg(\frac{m_V}{10\;\mathrm{MeV}}\bigg)^2 \bigg(\frac{100\;\mathrm{GeV}}{E}\bigg)~.
\end{equation}
This approximate form  is valid when $E \gg m_V$.
This exercise also shows that for $m_V > m_e$, the amplitude is enhanced for $x\approx 1$ as stressed in Ref.~\cite{Bjorken:2009mm}. 

The $q^2 \approx -Q^2_\mathrm{min}$ regime mentioned above is relevant because the photon propagator $\sim 1/q^2$ contributes another apparent singularity at small $\absQ^2$. 
This divergence is regulated in two ways: by the kinematic cut-off $Q^2_{\mathrm{min}}$, and by the atomic and nuclear form-factor $F(\absQ^2) \propto a^2 \absQ^2$ for $a^2 \absQ^2\lesssim 1$ 
with $1/a \approx 4.6Z^{1/3}\times 10^{-6}$ GeV (see \cref{G_el} and \cref{eq:atomic_para}). The latter effect is due to the nuclear charge being screened by the atomic electrons in this low momentum transfer limit. 
The kinematic cutoff is dominant when $m_V \gtrsim 3\;\mathrm{MeV} Z^{1/6} \sqrt{E/\mathrm{GeV}}$, which is the case for most of the parameter space we are interested in (especially if experimental detection places a ${\sim}$ GeV energy threshold on signal events, leading to a similar lower bound on the initial electron energy $E$). At large energies and small masses, however, the form factor suppression is relevant. 

To summarize, the bremsstrahlung amplitude has apparent singularities that are regulated by particle masses or atomic screening. This leads to a highly peaked differential cross-section that can be difficult to sample from. 
This is particularly challenging if the peaks occur in nontrivial surfaces in phase space (i.e., if they do not align with the integration variables). We find that the adaptive algorithm of \vegas~\cite{Lepage:1977sw,Lepage:2020tgj} is able to correctly integrate and sample from the partially integrated differential cross-section from Refs.~\cite{Tsai:1989vw, Liu:2017htz, Gninenko:2017yus}. We validate our implementation by comparing the total cross-section, energy and angular distributions of the emitted dark sector states to other calculations using \texttt{CalcHEP}~\cite{Belyaev:2012qa}, \madgraph~\cite{Alwall:2014hca} and the Weizs\"acker-Williams approximation~\cite{Liu:2017htz}. 

\begin{figure}
    \centering
    \includegraphics[width=0.47\textwidth]{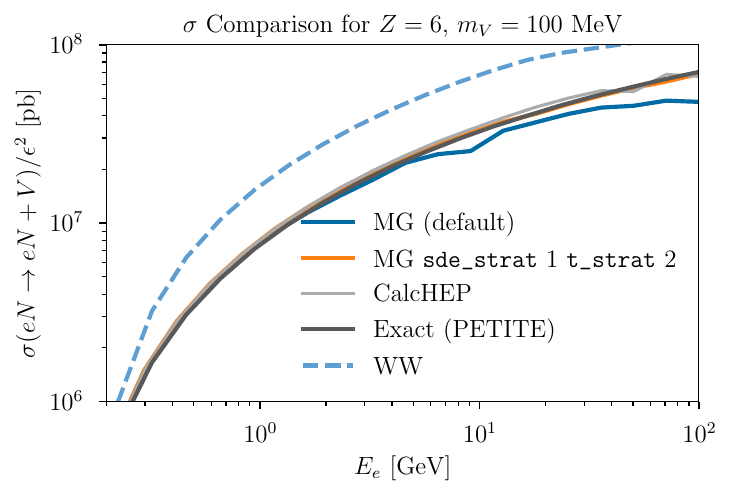}
    \caption{Total cross section for dark vector bremsstrahlung of electrons on a carbon nucleus for $m_V = 100$ MeV. We show \madgraph, \madgraph with settings that improve numerical stability, \calchep (which achieves good stability with default settings), the treatment in \PETITE, and the Weizsacker-Williams approximation commonly employed in the dark photon literature \cite{Bjorken:2009mm}. \label{fig:generator_cross_section_comparison}
    }
    
\end{figure}

\begin{figure}
    \centering
    \includegraphics[width=0.47\textwidth]{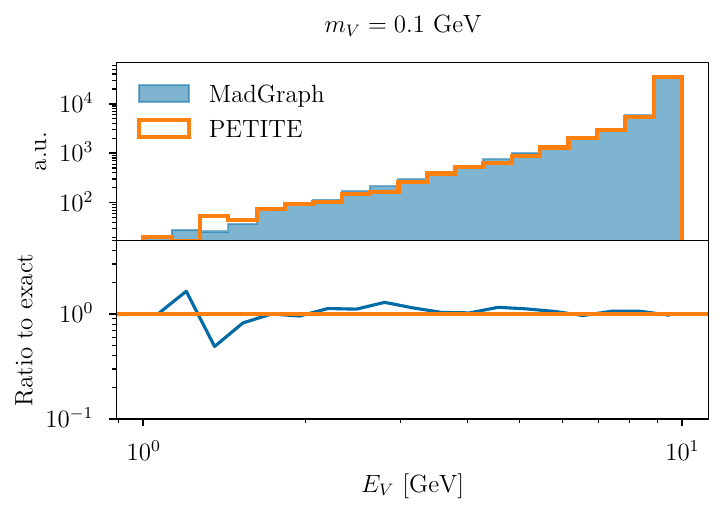}
    \includegraphics[width=0.47\textwidth]{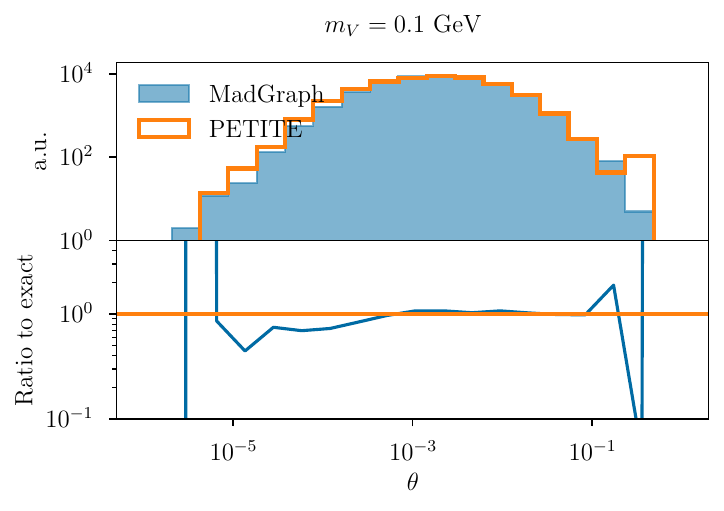}
    \caption{Energy (left) and angle (right) distributions of the emitted vector particle in $e^- + N \to e^- + N + V$ with an beam energy of $10$ GeV and $m_V =0.1$ GeV.}
    \label{fig:generator_distribution_comparisons}
\end{figure}

In \cref{fig:generator_cross_section_comparison} we compare the total cross section as a function of beam energy for a fixed $m_V = 0.1$ GeV and a carbon target evaluated using different methods. 
We see that our implementation is in agreement with existing MC tools, \texttt{CalcHEP} and \madgraph. 
We note that the default \madgraph settings (as of version \texttt{3.4.0}) are inappropriate for this particular process with no cuts on the emitted particles. 
This issue has been noted in, e.g., Refs.~\cite{Celentano:2020vtu,Forbes:2022bvo}. 
In the former, the atomic form factor was set to 0 for very low photon virtualities to avoid this problem.
The issue seems to be in the channel decomposition automatically performed by \madgraph \cite{Mattelaer:2021xdr}, which by default looks at the (singular) propagator $1/q^2$ instead of the (regular) combination $F(\absQ^2)/q^2$. 
We found that this results in a biased event sample, leading to inconsistent results in terms of cross-section and distributions. 
This problem is completely addressed by changing the single diagram enhancement strategy to \texttt{sde\_strat 1} and $t$ channel integration strategy to \texttt{t\_strat 2} (see Ref.~\cite{Mattelaer:2021xdr} for a detailed description of these options).
Both \texttt{CalcHEP} and our implementation simply use \vegas on the entire differential cross-section which avoids these complications. In \cref{fig:generator_distribution_comparisons} we show that angular and energy distributions of the emitted vector particles are in perfect agreement between \PETITE and \madgraph (with \texttt{sde\_strat 1} and \texttt{t\_strat 2}) within MC-statistical uncertainties.

\section{Dark Compton scattering and resonant annihilation \label{A:darkcompton}}
Dark Compton scattering involves an incident photon which scatters on an electron and produces an electron and dark vector, $e \gamma \rightarrow e V$. The differential cross section for dark Compton scattering is closely related to that for Compton scattering, \cref{eq:compton}:
\begin{equation}
    \dv{t}\sigma_{\rm Dark\ Comp.}=\kappaCoup^2 \dv{t}\sigma_{\rm Comp.} -\frac{4\kappaCoup^2\pi\alpha^2m_V^2}{(s-m_e^2)^4(u-m_e^2)^2}\left[s t u - m_e^2 m_V^2 (s+u) + m_e^2 \left(2 s u+m_e^2 \left(t-2m_e^2\right)\right) \right]~,
\end{equation}
where the Mandelstam variables in $\sigma_{\rm Comp.}$ are redefined and satisfy the relation $s+t+u = 2m_e^2 + m_V^2$.

In the $m_e\to 0$ limit, the differential cross sections for dark Compton scattering and annihilation $e^+ e^- \to \gamma V$
\begin{align}
    \dv{t} \sigma_{\rm Dark\;Comp.} &= \frac{2\kappaCoup^2 \pi \alpha^2}{s^2}\left[ \frac{ 2 m_V^2 (s+u) -2 m_V^4 -s^2-u^2}{s u}\right]~,\\
    \dv{t} \sigma_{\rm Dark\;Ann.} &= \frac{2 \kappaCoup^2 \pi \alpha^2}{s^2}\left[\frac{t u ( 2 m_V^4-2 m_V^2 (t+u)+t^2+u^2)}{(t-m_e^2)^2(u-m_e^2)^2}\right]~,
\end{align}
where the Mandelstam variables are $s = (p_\gamma + p_{e^-})^2$,  $u = (p_{e^-}-k_V)^2$  and $t = (p_\gamma-k_V)^2$ for Compton scattering. For the process of $e^+e^- \rightarrow \gamma V$ the Mandelstams are $s=(p_{e^+}+p_{e^-})^2$,  $u = (k_V-p_{e^-})^2$, $t = (k_\gamma - p_{e^-})^2$. For the dark annihilation differential cross section we retained $m_e$ in the denominator which regulates collinear divergences. Note that we do not use this expression in \PETITE; rather we implemented the results of a resummation of all soft and collinear SM emissions as described in \cref{A:radiative-return}.
Since the phase space of these processes is one-dimensional, sampling of the final state kinematics is fast. This is not so for dark bremsstrahlung, which is discussed in detail in \cref{A:darkbremm}.

\section{Positron annihilation to dark vectors \label{A:radiative-return} }
The importance of secondary positron annihilation for the production of dark sector states in thick targets was first pointed out in Ref.~\cite{Bjorken:1988as} in the context of axions and in Ref.~\cite{Marsicano:2018krp} for dark photons. At tree level, there are two processes $e^+ e^- \to V$ and $e^+ e^- \to \gamma V$ that can be called 
annihilation (termed ``resonant'' and ``non-resonant''). The distinction between them is not obvious since in the second process, the SM photon can be arbitrarily soft. The authors of Ref.~\cite{Marsicano:2018krp} avoid this issue by putting an (arbitrary) minimum energy cut on the SM photon. Additionally, the process $e^+ e^- \to \gamma V$ contains a collinear singularity (regulated by $m_e$) when the emitted photon is aligned with $e^\pm$; this leads to the presence of large logarithms in the rate, which make perturbative calculations worse. To address these issues, we consider an alternative approach to simulating these processes. 

The emission of an arbitrary number of soft or collinear photons can be accounted for by using electron and positron ``parton distribution functions'' (PDFs), $f_e(x, Q)$. The PDFs calculate the probability for an electron/positron to carry a fraction $x$ of longitudinal momentum of the beam particle when probed at scale $Q$. The largest annihilation rate occurs when the soft and collinear photon emissions bring the centre-of-mass energy of the colliding ``partons'' on resonance with the mass of the dark sector state: $x_+ x_- s = m_V^2$, where $x_\pm$ are the longitudinal momentum fractions of $e^\pm$. The cross-section for this radiative return is given by 
\begin{equation}
    \sigma_{rr}(s) = \left[\frac{4\pi^2 \alpha \varepsilon^2}{s}\beta_f \left(\frac{3-\beta_f^2}{2}\right)\right]
\int_{m_V^2/s}^{1} \frac{dx}{x} f_e(x,s) f_e\left(\frac{m_V^2}{xs},s\right),
\label{eq:rr_total_cross_section}
\end{equation}
where $\beta_f^2 = 1 - 4m_e^2/m_V^2$ and we used the narrow width approximation. 
The PDFs are calculated by solving the Gribov-Lipatov equation, which resums virtual, soft, and collinear emissions. A useful compilation of various approximations for $f_e(x)$ can be found in Ref.~\cite{Nicrosini:1986sm}. We will use the Kuraev-Fadin~\cite{Kuraev:1985hb} expression\footnote{As illustrated in Ref.~\cite{Greco:2016izi}, this PDF works best as $x\to 1$; we use it because of its simplicity.}
\begin{equation}
    f_e(x, s) = \frac{\beta}{16} \left((8+3\beta)(1-x)^{\beta/2 - 1}-4(1+x)\right)+\dots,
\end{equation}
where $\beta = (2\alpha/\pi)(\ln s/m^2_e - 1)$ and the ellipsis stands for higher order terms in $\beta$. 
Here we see the impact of the collinear singularities in factors of $\ln s/m_e^2$, and the soft divergence as $(1-x)^{1- \beta/2}$. For the typical GeV-scale energies $\log s/m_e^2$ is $\mathcal{O}(10)$. The resummation therefore has important effects both at large incoming particle energies (where these logarithms grow), and also near threshold $s\approx m_V^2$.
In particular, evaluating \cref{eq:rr_total_cross_section} as  $s\to m_V^2$, one finds that $\sigma_{rr}\propto (s - m_V^2)^{-1+\beta}$, compared to $(s - m_V^2)^{-1}$ in the tree-level approximation for $e^+e^- \to\gamma V$. 
In other words, the soft photon singularity becomes integrable after resummation. 
This is important when the cross-section is integrated with a distribution of incoming positron energies, as effectively happens in a EM shower in a beam dump. 
The resummed result gives a finite answer, while the tree-level cross-section would give infinity (without a minimum energy cut).

\begin{figure}[h]
    \centering
    \includegraphics[width=0.47\textwidth]{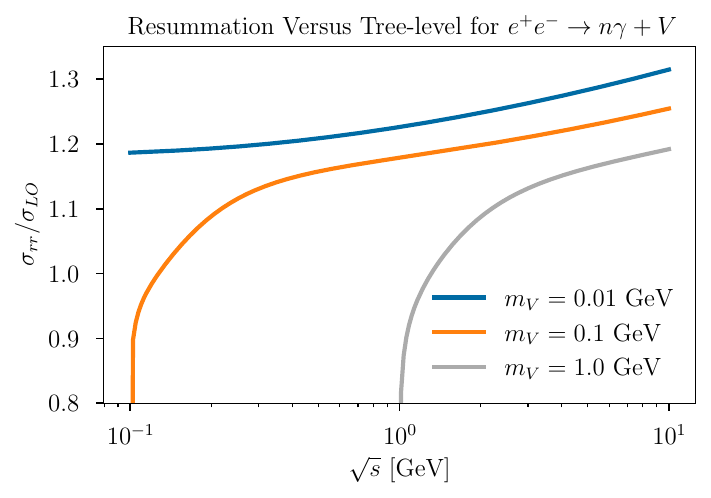}
    \caption{Ratio of the radiative return cross-section $\sigma_{rr}$ to the (leading order) tree-level annihilation cross-section $\sigma_{LO}$ for several dark vector masses, as a function of $\sqrt{s}$. The cross section is suppressed by an $O(1)$ fraction in the vicinity of the resonance because emission of soft-collinear photons take the parton below the resonance threshold. The cross section is enhanced for the same reason at higher $\sqrt{s}$, because radiative return allows the parton to {\it access} the resonance.  \label{fig:radiative_return_versus_tree_level}}
\end{figure}

In order to simulate the radiative return production of dark sector states, we sample the 
parton luminosity function (the integrand in \cref{eq:rr_total_cross_section}) to generate momentum fractions 
$x$ and $m_V^2/(x s)$ carried by leptons in the CM frame. 
The direction of the outgoing $V$ is taken to be in the same direction of the incoming $e^+$ and its energy is approximately $m_V^2/(2 x m_e)$.
While this adequately models the dominant part of the phase space (the one experiencing a collinear enhancement), one may be interested in events where a photon is emitted with a significant transverse momentum. 
Here the tree-level calculation would be appropriate since one can use it to explicitly generate the outgoing photon kinematics. These two parts of phase space can be merged by selecting cuts on the maximum $p_T = Q$ in the PDF calculation to correspond to the minimum $p_T$ in the tree-level calculation. 
For simplicity, we only focus on the collinearly-enhanced phase space in this work. In \cref{fig:radiative_return_versus_tree_level} we compare radiative return and tree-level cross section calculations. We find that radiative return modifies the cross section at the level of $10-20\%$. 

\pagebreak

\vfill 

\typeout{}
\bibliographystyle{JHEP}
\bibliography{refs}

\end{document}